\documentclass[%
 reprint,
nofootinbib,
 amsmath,amssymb,
 aps,
]{revtex4-2}

\usepackage{graphicx}
\usepackage{dcolumn}
\usepackage{bm}
\usepackage{hyperref}

\usepackage{commath}
\usepackage{mathrsfs}

\usepackage[dvipsnames]{xcolor}

\newcommand{\hypgeo}[2]{%
  \operatorname{%
    {\vphantom{\mathnormal{F}}}_{#1}%
    \kern-\scriptspace
    \mathnormal{F}_{#2}%
  }%
}

\DeclareMathOperator{\arccot}{arccot}
\DeclareMathOperator{\atantwo}{atan2}

\begin{document}

\preprint{APS/123-QED}

\title{Relativistic disks by Appell-ring convolutions}

\author{D. Kofroň}
 \email{k.kofron@gmail.com}
\author{P. Kotlařík}%
 \email{kotlarik.petr@gmail.com}
\author{O. Semerák}
 \email{oldrich.semerak@mff.cuni.cz}
\affiliation{
Institute of Theoretical Physics, Faculty of Mathematics and Physics,\\
Charles University,\\
V Hole\v{s}ovi\v{c}k\'{a}ch 2, 180\,00 Prague 8, Czech Republic}

\date{\today}

\begin{abstract}
    We present a new method for generating the gravitational field of thin disks within the Weyl class of static and axially symmetric spacetimes. Such a gravitational field is described by two metric functions: one satisfies the Laplace equation and represents the gravitational potential, while the other is determined by line integration. We show how to obtain analytic thin-disk solutions by convolving a certain weight function -- an Abel transformation of the physical surface-density profile -- with the Appell-ring potential. We thus re-derive several known thin-disk solutions while, in some cases, completing the metric by explicitly computing the second metric function. Additionally, we obtain the total gravitational field of several superpositions of a disk with the Schwarzschild black hole. While the superposition problem is simple (linear) for the potential, it is mostly not such for the second metric function. However, in particular cases, both metric functions of the superposition can be found explicitly. Finally, we discuss a simpler procedure which yields the potentials of power-law-density disks we studied recently.
\end{abstract}

\maketitle


\section{Introduction} \label{sec:intro}
When matter with modest energy has enough angular momentum, it often forms disk-like structures due to the mutual competition between gravitational and centrifugal forces. Such structures occur in many astrophysical phenomena from galaxy scales down to accretion disks around stellar-mass objects or even planets. Disks are crucial in high-energy processes which drive active galactic nuclei, X-ray binaries, and gamma-ray bursts. They are important in observations of black-hole silhouettes, and they may affect the future gravitational-wave images of compact objects.

In order to describe the disk gravitational field in general relativity (GR), one has to either impose some reasonable simplifications to tackle the problem analytically, or to solve the Einstein equations numerically. In this work, we follow the first approach -- we propose a generic procedure which in some static and axially symmetric cases provides the full spacetime metric explicitly and in closed form.

For a physical system in a stationary equilibrium, if solenoidal-type motions are not important, it is natural to assume that its spacetime is circular, namely that it is stationary, axially symmetric, and orthogonally transitive (possessing global meridional planes). If the overall net rotation can be neglected, or if it is compensated due to counter-rotating mass currents, one can even consider the spacetime static. In such a case, if the condition $T^\rho_\rho+T^z_z=0$ on the energy-momentum tensor holds (as e.g. in a vacuum or dust), the metric can be written in the Weyl form
\begin{equation}
    \dif s^2 = - e^{2\nu} \dif t^2 + \rho^2 e^{-2\nu} \dif\phi^2 + e^{2\lambda - 2\nu} (\dif\rho^2 + \dif z^2) \,,
\end{equation}
where the Weyl-type cylindrical coordinates $(t,\rho,z,\phi)$ are adapted to the symmetries -- the metric is independent of the time $t$ and of the azimuthal angle $\phi$, while $\rho$ and $z$ cover the meridional planes in an isotropic way. In vacuum regions, the two metric functions $\nu(\rho,z)$ and $\lambda(\rho,z)$ satisfy the Einstein equations
\begin{align}
    \Delta \nu &= 0 \label{eq:LaplaceEquation} \,,\\ 
    \lambda_{,\rho} &= \rho ( \nu_{,\rho}^2 - \nu_{,z}^2) \,, \qquad   \lambda_{,z} = 2\rho \nu_{,\rho} \nu_{,z} \label{eq:lambda} \,.
\end{align}
where $\Delta$ is the standard Laplace operator in cylindrical coordinates $(\rho,z,\phi)$ (with the $\phi$ term missing, however, thanks to the axial symmetry). Hence, just like in Newtonian gravity or in electrostatics, the field is described by solutions to the Laplace equation. However, in contrast to the former, in GR there also appears the second function $\lambda$ which can deviate the result from the Newtonian expectations significantly. For instance, the Bach-Weyl ring is the natural analogue of the homogeneous thin ring from Newton's theory (they share the same form of the potential $\nu$), but the meridional plane is very much deformed close to the ring -- the space is not locally cylindrical around the ring, but rather strongly anisotropic~\cite{semerak2016a}.

We focus on infinitesimally thin disks here. Such a case finds applications where the typical cross thickness of the disk is negligible compared to other length scales. As relativists, we are mostly interested in disks formed in accretion in strong gravitational fields, e.g. that of a black hole or a neutron star. When the disk mass is much smaller than the mass of the central body, the accreting matter is usually treated as a test, non-gravitating one. Yet the field equations of Newton's as well as Einstein's theory involve mass {\em density}, so even if the total mass of the disk is not large, its gravity may still have a significant effect, including that on the disk's own structure, evolution and stability \cite{karas2004}. The disk gravitation has also been shown to induce subtle but possibly measurable effects on quasi-normal modes of black holes \cite{chen2023} or on extreme-mass-ratio inspirals \cite{polcar2022}, thus affecting the gravitational waves generated by the black-hole--disk systems; similarly, it can alter the appearance of the black-hole silhouettes \cite{cunha2020}. Having an analytical model which includes the gravitational contribution of the disk may thus help to understand such phenomena.

Newtonian-density profiles\footnote
{The quantity $\sigma$ satisfies the Poisson equation $\Delta\nu=4\pi\sigma(\rho,z)$, so it is the analogue of the Newtonian matter density.}
of static and axisymmetric razor-thin disks are proportional to the delta distribution in the $z$ coordinate,
\begin{equation}
    \sigma (\rho, z) \equiv \sigma(\rho) \delta(z) \,.
\end{equation}
The spacetime is assumed to be reflection symmetric with respect to the disk plane ($z\!=\!0$), regular and asymptotically flat. The potential $\nu$ as well as the second metric function $\lambda$ thus should be finite everywhere and vanish at spatial infinity ($\sqrt{\rho^2+z^2}\rightarrow\infty$).

The disk potential $\nu$ is given by the Poisson integral
\begin{align}
    &\nu (\rho, z) = \nonumber\\ &-4 \int_0^\infty \frac{\sigma (\rho^\prime) \rho^\prime}{\sqrt{(\rho + \rho^\prime)^2 + z^2}} K \left( \frac{2\sqrt{\rho\rho^\prime}}{\sqrt{(\rho + \rho^\prime)^2 + z^2}} \right) \dif \rho^\prime \;, \;\; \label{eq:PoissonIntegral}
\end{align}
where $K(k)$ is the complete elliptic integral of the first kind, with $k$ as its modulus. Since it is usually impossible to compute the integral  (\ref{eq:PoissonIntegral}) analytically, several other methods have been developed in the literature. Inspired by Kuzmin's trick (see Sec.~\ref{sec:solvingEinsteinEquations}), we propose a new way to solve the problem by integrating, over the Abel transformation of the source density profile, the ``Green function'' given by the potential due to the Appell ring. We illustrate the procedure on several known thin-disk solutions while completing them by also finding the second metric function $\lambda$ explicitly in some cases.

The plan of the paper is as follows. In Sec.~\ref{sec:solvingEinsteinEquations} we review some techniques for solving the Weyl-Einstein equations (\ref{eq:LaplaceEquation}), (\ref{eq:lambda}). In Sec.~\ref{sec:StaticThindisks}, we list several explicit disk solutions, specifically the Morgan-Morgan disks and their inverted counterparts, the disks with polynomial and power-law densities, the Kuzmin-Toomre and the Vogt-Letelier disks; we mention in particular the case of the Appell disk/ring which will serve as an elementary solution in our method. The method itself is introduced in Sec.~\ref{sec:convolution}. Then in Sec.~\ref{sec:particularSolutions} we demonstrate the applicability of the approach on the above particular disk solutions. In addition, we generalize the Morgan-Morgan disks and superpose them with a Schwarzschild black hole, while also deriving the second metric function $\lambda$ for the complete field. Finally, we revisit the case of power-law-density disks \cite{kotlarik2022} and describe a simpler procedure for generating their potential. Concluding remarks are added in Sec.~\ref{sec:conclusions}.

We use geometrized units in which the speed of light $c$ and the gravitational constant $G$ equal unity. The metric signature is ($-$$+$$+$$+$). The mass of the black hole is everywhere denoted by $M$ and the mass of the disk by~${\cal M}$.

\section{Thin-disk solutions to Einstein equations} \label{sec:solvingEinsteinEquations}
Let us outline some techniques for solving the equations (\ref{eq:LaplaceEquation}), or (\ref{eq:PoissonIntegral}), and (\ref{eq:lambda}) for a thin-disk distribution of matter. We focus on those which provide the solutions considered in the next section.

\subsection{The potential of thin disks}

One way to tackle the axisymmetric problem is to perform the integration (\ref{eq:PoissonIntegral}) along the symmetry axis $\rho\!=\!0$ where $K(k)$ reduces to $\pi/2$. If the result can be expanded in a power series in $z$, the solution at the generic location is given by a similar series supplemented by Legendre polynomials  \cite{jackson1999}. Such a series typically does not converge very well \cite{semerak2004}, though, in the case of the Morgan-Morgan family of disks, it performs better (see Sec. \ref{sec:MorganMorgan} for more details). 

A different approach was used by Conway \cite{conway2000} and we have applied it in \cite{kotlarik2022} recently. The trick is to rewrite the integral (\ref{eq:PoissonIntegral}) in terms of the Laplace transform of a product of the zero-order Bessel functions,
\begin{equation}\label{eq:conwayIntegral}
    \nu (\rho, z) = -2 \pi \int_0^\infty \int_0^\infty \rho^\prime \sigma (\rho^\prime) J_0 (s \rho^\prime) J_0 (s \rho) e^{-s |z|} \dif s \dif \rho^\prime \,,
\end{equation}
where $s$ is an auxiliary real parameter of the dimension of inverse length. The double integration may look complicated, but it has proved quite efficient, specifically in the case of the polynomial and power-law density disks where it even yields closed-form formulae \cite{kotlarik2022} (see Sec. \ref{sec:Polynomialdisks}). 

Yet another useful approach is due to Kuzmin \cite{kuzmin1956}, later elaborated extensively by Evans \& de Zeeuw \cite{evans1992} in the Newtonian context and by Bi\v{c}\'ak et al. \cite{bicak1993,bicak1993a} in GR. It involves considering a line mass distribution along the negative-$z$ half of the symmetry axis, cutting the resulting gravitational field along the $z\!=\!0$ plane, and copying (reflecting) the result obtained for the positive-$z$ half-space below that plane. The resulting potential reads
\begin{equation}\label{eq:kuzminIntegral}
    \nu (\rho, z) = - \int_{0}^{\infty} \frac{W(b) \dif b}{\sqrt{\rho^2 + (|z| + b)^2}} \;,
\end{equation}
with the ``weight function'' $W(b)$ describing the mass distribution, and the surface density induced by the metric-gradient jump in the equatorial plane reads
\begin{equation}
    \sigma(\rho) = \frac{1}{2\pi} \int_{0}^{\infty} \frac{W(b) b}{(\rho^2 + b^2)^{3/2}} \dif b \,.
\end{equation}

Note that Vieira \cite{vieira2020} employed a similar procedure in order to compute the potential of a black hole surrounded by a disk. His mass distribution was a chain of Schwarzschild black holes (rods) located on the axis, and he performed the cut and reflection in the middle of the positive-$z$ one. The Kuzmin procedure can also be generalized to a stationary case if glueing along suitable hypersurfaces \cite{ledvinka2019}. 

The potentials found by any of the above methods can be further extended by superpositions thanks to the linearity of the Laplace equation and by inversion with respect to a sphere of a certain prescribed radius (this so-called Kelvin transformation again yields a solution). The inversion with respect to the radius $b$ changes the position as
\begin{equation}
    \rho \longrightarrow \frac{b^2 \rho}{\rho^2 + z^2} \,, \quad z \longrightarrow \frac{b^2 z}{\rho^2 + z^2} \,, \label{eq:kelvinCoor}
\end{equation}
and the original potential $\nu(\rho,z)$ transforms to
\begin{equation}
   \mathcal{K} \nu(\rho, z) \equiv \frac{b}{\sqrt{\rho^2 + z^2}}\, \nu \left( \frac{b^2 \rho}{\rho^2 + z^2}, \frac{b^2 z}{\rho^2 + z^2} \right) \,. \label{eq:kelvinPot}
\end{equation}
It corresponds to a different (``inverted'') density profile, which in the case of a thin disk reads
\begin{equation}
    \sigma (\rho) \longrightarrow \frac{b^3}{\rho^3}  \sigma(b^2/\rho) \,.
\end{equation}

Besides the Weyl coordinates, we will also use the oblate ones, $\zeta\in(0,\infty)$ and $\xi\in(-1,+1)$, defined by
\begin{equation}
    \rho^2 = b^2 (1 + \zeta^2)(1 - \xi^2) \,, \qquad z = b \zeta \xi \,.
    \label{eq:CylToOsph}
\end{equation}
The inverse relations read
\begin{align}
    \zeta &= \frac{\sqrt{2} |z|}{\sqrt{\sqrt{(\rho^2 - b^2 + z^2)^2 + 4b^2z^2} - (\rho^2 - b^2 + z^2)}} \,,\qquad \\
    \xi &= \frac{z}{b \zeta} \,,
\end{align}
and the Kelvin transformation works as
\begin{align}
    &\zeta \longrightarrow \frac{\xi}{\sqrt{\zeta^2 + 1 - \xi^2}} \,, \qquad \xi \longrightarrow \frac{\zeta}{\sqrt{\zeta^2 + 1 - \xi^2}} \,. 
    \label{eq:kelvinOsphCoor} \\
    &\mathcal{K} \nu(\zeta, \xi) = \nonumber\\
    &\frac{1}{\sqrt{\zeta^2 + 1 - \xi^2}}\, \nu \left(\frac{\xi}{\sqrt{\zeta^2 + 1 - \xi^2}}, \frac{\zeta}{\sqrt{\zeta^2 + 1 - \xi^2}} \right) \,.
    \label{eq:kelvinOsphPot}
\end{align}

\subsection{The function $\lambda$ and superposition of multiple sources} \label{subsec:solvingLambda}

For $\lambda$, there is no other option usually than to integrate equation (\ref{eq:lambda}) directly, although the reflecting method works for the whole metric, thus also for $\lambda$. In a vacuum, the main requirement on $\lambda$ is to vanish on the axis (the ``elementary flatness'' requirement), plus we assume $\lambda\!=\!0$ at spatial infinity (asymptotic flatness).

Thanks to the linearity of the Laplace equation, the potentials of multiple sources just add, e.g. $\nu=\nu_1+\nu_2$, yet still, the non-linearity of the Einstein equations shows itself in the second metric function $\lambda$. Denoting the pure individual contributions as $\lambda_1$ and $\lambda_2$ (they satisfy (\ref{eq:lambda}) with $\nu_1$ and $\nu_2$ respectively), the total $\lambda$ (satisfying (\ref{eq:lambda}) with $\nu_1+\nu_2$) comes out as 
\begin{equation}
    \lambda = \lambda_1 + \lambda_2 + \lambda_\text{int} \,,
\end{equation}
where the ``interaction'' part $\lambda_{\rm int}$ has gradient
\begin{align}
    \lambda_{\text{int}, \rho} &= 2 \rho (\nu_{1, \rho} \nu_{2, \rho} - \nu_{1, z} \nu_{2, z}) \,, \label{eq:lambdaInt1} \\
    \lambda_{\text{int}, z} &= 2 \rho (\nu_{1, \rho} \nu_{2, z} + \nu_{1, z} \nu_{2, \rho}) \,. \label{eq:lambdaInt2}
\end{align}
In particular, if ``the first'' of the sources is the Schwarzschild black hole (of mass $M$), the corresponding metric functions read
\begin{align}
    \nu_1 \equiv \nu_\text{Schw} &= \frac{1}{2}\ln\left(\frac{R_++R_--2M}{R_++R_-+2M}\right)\,, \\
    \lambda_1 \equiv \lambda_\text{Schw} &= \frac{1}{2}\ln\left(\frac{(R_++R_-)^2-4M^2}{4R_+R_-}\right) \,,
\end{align}
where
\begin{equation}
    R_\pm = \sqrt{\rho^2+(z \mp M)^2}\,.
    \label{eq:RpRmDef}
\end{equation}

\section{Specific thin disks}\label{sec:StaticThindisks}
In this section, we review several solutions for thin disks which are empty in the central region (or which can be made such by inversion), being thus suitable for the superposition with a central source.

\subsection{Morgan-Morgan disks and their inversion}\label{sec:MorganMorgan}

A class of general relativistic thin disks with the Newtonian density profile
\begin{equation}
    \sigma^{(n)}_\text{MM} (\rho \leq b) = \frac{(2n + 1) \mathcal{M}}{2\pi b^2} \left( 1 - \frac{\rho^2}{b^2} \right)^{n-1/2}
    \label{eq:sigmaMM}
\end{equation}
was proposed by Morgan \& Morgan already in 1969~\cite{morgan1969}. These disks are finite, stretching between $\rho \in [0, b]$ and having a finite total mass $\mathcal{M}$. Their field is described by the potential \cite{semerak2003}
\begin{equation}
    \nu_\text{MM}^{(n)} = - \frac{\mathcal{M}}{b} (2n + 1)! \sum_{j=0}^{n} C_{2j}^{(n)} i Q_{2j}(i \zeta) P_{2j} (\xi) \;,
    \label{eq:nuMM}
\end{equation}
where $P_{2j}$ and $Q_{2j}$ are the Legendre functions of the first and second kind and the coefficients read
\begin{equation}
    C_{2j}^{(n)} = \frac{(-1)^j(4j + 1)(2j)!(n+j)!}{(j!)^2(n-j)!(2n + 2j + 1)!} \,.
\end{equation}
It is also possible to obtain the second metric function $\lambda$ case by case integrating the Einstein equations (\ref{eq:lambda}) in the oblate spherical coordinates (see Sec. \ref{sec:hMM}).

If interested in infinite disks with a central empty region, an inversion with respect to the outer rim~$b$, i.e. the Kelvin transformation (\ref{eq:kelvinPot}), can be applied. The resulting field corresponds to the disks with the Newtonian density profiles
\begin{equation}
    \sigma_\text{iMM}^{(n)} (\rho \geq b) = \frac{2^{2n}(n!)^2 \mathcal{M}b}{(2n)! \pi^2 \rho^3} \left( 1 - \frac{b^2}{\rho^2} \right)^{n-1/2} \,.
\end{equation}
They stretch from $\rho\!=\!b$ to radial infinity, yet their total mass $\mathcal{M}$ remains finite. There is no explicit result for $\lambda$ in the literature. The inverted Morgan-Morgan disks were first considered and superposed with the Schwarzschild black hole by Lemos \& Letelier \cite{lemos1994}.

\subsection{Kuzmin-Toomre and Vogt-Letelier disks}

When the weight function $W(b)$ in (\ref{eq:kuzminIntegral}) is proportional to a certain sum of delta distributions and their derivatives, the associated potential describes the gravitational field of a particular family of thin disks studied by Kuzmin \cite{kuzmin1956} and Toomre \cite{toomre1963},
\begin{equation}
    \nu_\text{KT}^{(n)} = -\frac{\mathcal{M}}{(2n - 1)!!} \sum_{k=0}^n \frac{(2n-k)!}{2^{n-k}(n-k)!} \frac{b^k}{r_b^{k+1}} P_k\left( |\cos\theta_b| \right) \,,
    \label{eq:KTpotential}
\end{equation}
where we have denoted
\begin{equation}
    r_b^2 \equiv \rho^2 + (|z| + b)^2 \,, \qquad |\cos\theta_b| \equiv \frac{|z| + b}{r_b} \,,
\end{equation} 
and $P_k$ are Legendre polynomials. The corresponding Newtonian density profiles read
\begin{equation}
    \sigma_\text{KT}^{(n)} = \frac{(2n+1)b^{2n + 1}}{2\pi} \frac{\mathcal{M}}{(\rho^2 + b^2)^{n + 3/2}}  \,.
    \label{eq:sigmaKT}
\end{equation}
The second metric function $\lambda$ was found by Bičák et al. in \cite{bicak1993}. 

Taking a special superposition of expressions (\ref{eq:KTpotential}), Vogt \& Letelier \cite{vogt2009} obtained another potential-density pairs
\begin{align}
    \nu_\text{VL}^{(m, n)} &=W^{(m,n)} \sum_{k=0}^n (-1)^k \binom{n}{k} \frac{\nu^{(m + k)}_\text{KT}}{2m+2k+ 1} \,,
    \label{eq:nuVLoriginal}\\
    \sigma_\text{VL}^{(m, n)} &= W^{(m, n)} \frac{\mathcal{M} b^{2m+1}}{2\pi} \frac{\rho^{2n}}{(\rho^2 + b^2)^{m + n+ 3/2}} \,,
    \label{eq:VLdensity}
\end{align}
where $W^{(m, n)}$ is a normalization factor ensuring that $\mathcal{M}$ remains the total mass of the disk. Recently, in \cite{kotlarik2022a}, we derived the second metric function $\lambda$ and, using the fact that these disks are empty on the axis, we superposed them with the Schwarzschild black hole and provided both metric functions explicitly for the total spacetime.

\subsection{Polynomial and power-law density disks}\label{sec:Polynomialdisks}

In \cite{kotlarik2022}, we presented solid finite disks of polynomial density profiles, infinite annular disks of power-law density and finite annular disks with a bump-like density,
\begin{align}
    \sigma^{(m, 2l)}_\text{pol} (\rho \leq b) &= \binom{m + \frac{1}{l}}{m} \frac{\mathcal{M}}{\pi b^2} \left( 1 - \frac{\rho^{2l}}{b^{2l}} \right)^m \,, \\
    \sigma^{(m, 2l)}_\text{pl} (\rho \geq b) &= \binom{m + \frac{1}{2l}}{m}  \frac{\mathcal{M} b}{2\pi \rho^3} \left( 1 - \frac{b^{2l}}{\rho^{2l}} \right)^m \,, \\
    \sigma_\text{bump}^{(L)} (b_\text{in} \leq \rho & \leq b_\text{out}) = - W_0 + \sum_{j = 0}^{L} (-1)^j \frac{W_{-3-2j}}{\rho^{3 + 2j}} \,,
\end{align}
where $W_i$ are constant coefficients. The construction employed the special properties of the Laplace equation, i.e., we integrated (\ref{eq:conwayIntegral}) for elementary density terms $\sigma_\text{pol}^{(2l)} = \rho^{2l}$, applied Kelvin transformation (\ref{eq:kelvinCoor}) and (\ref{eq:kelvinPot}) to get the negative powers $\rho^{-3-2l}$, and finally superposed the corresponding elementary potentials with appropriate weights. The resulting potentials were given in closed form in terms of complete elliptic integrals. The second metric function $\lambda$ has not been found explicitly.

\subsection{Appell ring}

Already in 1887, Appell \cite{appell1887} obtained, within electrostatics, a ring-like solution to the Laplace equation by putting a charged point particle (massive particle in our case, $\mathcal{M}$) on the complex extension of the $z$ axis. The corresponding complex potential reads
\begin{equation}
    \nu_\text{App} = -\frac{\mathcal{M}}{\sqrt{\rho^2 + (z - \text{i} b)^2}} \;,
\end{equation}
with a point mass located at $(\rho = 0, z = \text{i}b)$, where $b$ is real and of the dimension of length. Taking the real part, we have
\begin{align}
    \nu_\text{App} &= -\frac{\mathcal{M}}{\sqrt{2} \, \Xi} \sqrt{\Xi + \rho^2 + z^2 - b^2} \,, \label{eq:appellPotential}\\
    \lambda_\text{App} &= \frac{\mathcal{M}^2}{8b^2} \left[ 1 - \frac{\rho^2 + z^2 + b^2}{\Xi} - \frac{2b^2\rho^2 (\Xi^2 - 8z^2b^2)}{\Xi^4} \right] \,,
\end{align}
where 
\begin{equation}
    \Xi = \sqrt{(\rho^2 - b^2 + z^2)^2 + 4b^2z^2} \,.
\end{equation}
A thorough investigation of the whole broader family of Appell rings in GR was performed by Gleisler \& Pullin \cite{gleiser1989}. Two interpretations of the simplest solution are at hand. It is either a disk of negative surface mass density diverging towards $-\infty$ at the rim $(\rho = b, z=0)$, jumping to $+\infty$ there, so that the total mass $\mathcal{M}$ is positive and finite, or, it is a singular ring of mass $\mathcal{M}$, through which the spacetime is analytically extended to a second asymptotically flat region. Semerák et al. \cite{semerak1999} pointed out that the gravitational field is somewhat similar to the Kerr solution, although no dragging effects are present as the spacetime is static, and there is also no horizon. Independently of the interpretation, we use the Appell potential to generate physical thin-disk solutions similarly as Kuzmin did in \cite{kuzmin1956} for the point mass on the (real) axis.

\section{Convolution of Appell rings}\label{sec:convolution}

\subsection{The potential $\nu$}

Let us combine the Kuzmin-inspired approach (\ref{eq:kuzminIntegral}) and Appell's trick (\ref{eq:appellPotential}), i.e., integrate a line matter distribution on the complex extension of the $z$ axis described by the real \textit{weight function} $f(w)$ and mirror the positive half of the $z$ axis to the negative half. The respective potential thus reads
\begin{equation}
    \nu_f (\rho, z) = -\int_0^\infty \frac{f(w) \dif w}{\sqrt{\rho^2 + (|z| + \text{i} w)^2}} \,,
    \label{eq:appellconvolution}
\end{equation}
Note that $\nu_f$ is a complex function at this moment. Its real part corresponds to some matter distribution in the equatorial plane, while its imaginary part corresponds to a dipole. Indeed, in the oblate spheroidal coordinates (\ref{eq:CylToOsph}), for $\zeta \gg 1$, we have
\begin{equation}
    \frac{f(w) \dif w}{\sqrt{\rho^2 + (|z| + \text{i} w)^2}} \approx -\frac{f(w)}{b \zeta} + \frac{\text{i} w \xi f(w)}{b^2 \zeta^2} + \mathcal{O}(\zeta^{-3}),
\end{equation}
so $\int_0^\infty f(w) \dif w$ may be interpreted as the total mass. The imaginary part of (\ref{eq:appellconvolution}) we will not consider any further, because it brings the non-physical dipole term.

The integral (\ref{eq:appellconvolution}) can be rewritten as 
\begin{equation} \label{eq:appellconvolutionComplexIntegral}
    \nu_f (\rho, z) = - \text{i} \int_\gamma f\left( \frac{1}{2\text{i}}  \frac{\rho^2 + z^2 - t^2}{t - z} \right) \frac{\dif t}{t - z}
\end{equation}
if holomorphically extending the function $f$ to a complex plane, with the integration contour
\begin{equation}
    \gamma(w): \; t = \sqrt{\rho^2 + (z + \text{i} w)^2} - \text{i} w \,, \qquad w \in [0, \infty) \,.
\end{equation}
The integrand is holomorphic in the neighbourhood of the contour $\gamma$, so there exists an antiderivative $F$ to (\ref{eq:appellconvolutionComplexIntegral}) such that
\begin{equation}
    \nu_f = F\left[\gamma(w = \infty)\right] - F\left[\gamma(w = 0)\right] \,.
\end{equation} 

The corresponding (yet complex) Newtonian density profile reads
\begin{align}
    \sigma_f (\rho) &= \frac{1}{2\pi} \lim_{z \rightarrow 0^+} \dpd{\nu_f}{z} \\
    &= \frac{1}{2\pi}  \lim_{z \rightarrow 0^+}\int_0^\infty \frac{(z + \text{i} w) }{\left[\rho^2 + (z + \text{i} w)^2\right]^{3/2}} f(w) \dif w \,.
\end{align}
Assuming that $f(w)$ is continuous, bounded, and vanishes on the boundary, integration by parts yields
\begin{align}
    \sigma_f (\rho) &=  - \frac{\text{i}}{2\pi} \lim_{z \rightarrow 0^+}\int_0^\infty \dpd{f}{w} \frac{\dif w}{\sqrt{(z + \text{i} w)^2 + \rho^2}} \\
    & = - \frac{\text{i}}{2\pi} \int_0^\rho \dpd{f}{w} \frac{\dif w}{\sqrt{\rho^2 - w^2}} - \nonumber\\
    & \qquad\qquad\qquad\quad -\frac{1}{2\pi} \int_\rho^\infty \dpd{f}{w} \frac{\dif w}{\sqrt{w^2 - \rho^2}} \,,
\end{align}
where we took the limit $z \rightarrow 0^+$ and split the integral into two parts. The principal branch of natural logarithm is used, i.e. $\log{(-1)} = \text{i}\pi$, so $\sqrt{-1} \equiv e^{1/2\log(-1)}  = \text{i}$. Both integrands are real, thus only the second part contributes to the actual (physical) part of the Newtonian density,
\begin{equation}
    \sigma_f (\rho) =  -\frac{1}{2\pi}\int_\rho^\infty \dpd{f}{w} \frac{\dif w}{\sqrt{w^2 - \rho^2}} \,,
    \label{eq:sigmaf}
\end{equation}
which is the Abel transformation of the weight function~$f(w)$. Its inverse reads
\begin{equation}
    f(w) = 4 \int_w^\infty \frac{\sigma_f(\rho)}{\sqrt{\rho^2 - w^2}} \rho \dif \rho \,.
    \label{eq:weightFromDensity}
\end{equation}
Hence, rather than solving the challenging integral (\ref{eq:PoissonIntegral}), we can find the weight function $f(w)$ of the desired Newtonian density using (\ref{eq:weightFromDensity}) and evaluate a simpler integration (\ref{eq:appellconvolution}) to obtain the gravitational potential.

\subsection{The second metric function $\lambda$}

With the potential (\ref{eq:appellconvolution}) at hand, we can look for $\lambda$. Equations (\ref{eq:lambda}) are quadratic in $\nu$, thus working with the complex potential (\ref{eq:appellconvolution}) and taking the real part later would not yield the correct $\lambda$-counterpart of the real part of (\ref{eq:appellconvolution}): one has to solve (\ref{eq:lambda}) for the real part of (\ref{eq:appellconvolution}) directly. Similarly to \cite{bicak1993a}, we can express $\lambda$ as an integral over the individual $\nu$ pairs
\begin{equation}
    \lambda_f = \int_{0}^{\infty} \int_{0}^{\infty} f(w_1) f(w_2) \mathcal{F} (w_1, w_2) \dif w_1 \dif w_2 \,,
    \label{eq:lambdaDoubleInt}
\end{equation}
where $\mathcal{F}$ is the cross term between the individual real parts of the Appell potentials
\begin{equation}
    \nu_j = \frac{1}{2} \left[ \nu_\text{App} (w_j) + \overline{\nu}_\text{App}(w_j)\right] 
    \quad (j=1,2) \,,
\end{equation}
satisfying the same equations as the interaction part of $\lambda$, (\ref{eq:lambdaInt1}) and (\ref{eq:lambdaInt2}), with $\nu_j$. One finds that in general
\begin{equation}
    \mathcal{F} (w_1, w_2) = 2\frac{(w_1^2 + w_2^2)}{(w_1^2 - w_2^2)^2} - \frac{1}{2} \left[ \mathcal{H}(w_1, w_2) + \overline{\mathcal{H}}(w_1, w_2) \right] \,,
\end{equation}
with
\begin{align}
    \mathcal{H}(w_1, w_2) &\equiv h(w_1, w_2) + h(w_1, -w_2) \,, \nonumber\\
    h(w_1, w_2) &\equiv \frac{\rho^2 + (z + \text{i} w_1)(z + \text{i} w_2)}{(w_1-w_2)^2} \frac{1}{\mathscr{R}_1 \mathscr{R}_2} \,, \label{eq:crossTermh}
\end{align}
and with the notation
\begin{equation}
    \mathscr{R}_j^2 = \rho^2 + (z + \text{i} w_j)^2 \,.
\end{equation}
The constant term $2(w_1^2 + w_2^2)/(w_1^2 - w_2^2)^2$ appears due to the regularity condition on the axis, where we require $\lim_{\rho \rightarrow 0^+} \lambda = 0$.

We can now repeat the procedure for the Kelvin-inverted potential $\mathcal{K} \nu_f$. Introducing
\begin{equation}
    \mathcal{K} \mathcal{H}(w_1, w_2) \equiv \frac{w_1 w_2}{b^4} \left[ - \mathcal{K} h(w_1, w_2)  + \mathcal{K} h(w_1, -w_2)  \right]\,,
\end{equation}
the cross term in this case reads
\begin{align}
    \mathcal{F} (w_1, w_2) &= -\frac{4 w_1^2 w_2^2}{b^4 (w_1^2 - w_2^2)^2}  \nonumber\\
    & \qquad  - \frac{1}{2} \left[ \mathcal{K}\mathcal{H}(w_1, w_2) + \mathcal{K}\overline{\mathcal{H}}(w_1, w_2) \right] \,.
\end{align}
The term $\mathcal{K}h$ is the $h$ from (\ref{eq:crossTermh}) Kelvin-transformed according to (\ref{eq:kelvinCoor}).

The procedure works generally, yet in specific situations, it may be easier to solve the equations (\ref{eq:lambda}) directly, usually after transforming to some appropriate coordinates. For instance, in the oblate spheroidal coordinates $(\zeta, \xi)$, the equations for $\lambda$ read
\begin{align}
    \frac{\zeta^2 + \xi^2}{\xi^2 - 1} \lambda_{,\zeta} &=  -\zeta (\xi^2 - 1) \nu_{,\xi}^2 - \zeta (\zeta^2 + 1) \nu_{,\zeta}^2 + \nonumber\\
    & \qquad\qquad\qquad\qquad + 2\xi (\zeta^2 + 1) \nu_{,\zeta}\nu_{,\xi} \,, \label{eq:lambdaZeta}\\
    \frac{\zeta^2 + \xi^2}{\zeta^2 + 1} \lambda_{,\xi} &=  \xi (\zeta^2 + 1) \nu_{,\zeta}^2 + \xi (\xi^2 - 1) \nu_{,\xi}^2 - \nonumber\\
    & \qquad\qquad\qquad\qquad - 2 \zeta (\xi^2 - 1) \nu_{,\zeta} \nu_{,\xi} \,. \label{eq:lambdaXi}
\end{align}

\subsection{Superposition with a black hole}

When interested in a superposition of the disk with a black hole of mass $M$, the interaction part $\lambda_\text{int}$ is also necessary. Note that the equations (\ref{eq:lambdaInt1}) and (\ref{eq:lambdaInt2}) for $\nu_1 = \nu_\text{Schw}$ and $\nu_2 = \nu_f$ are linear in the disk contribution. Thus, the situation is in fact simpler since we can work with the complex potential $\nu_f$ and take the real part only at the end. One finds
\begin{multline}
    \lambda_\text{int} = \int_0^\infty \left\{ \frac{2Mf(w)}{w^2 + M^2} - \right.\\ 
    \left. \left[ \frac{R_+(\rho, z)}{M + \text{i} w} + \frac{R_-(\rho, z)}{M - \text{i} w} \right] \frac{f(w)}{\sqrt{\rho^2 + (z + \text{i} w)^2}} \right\} \dif w \,,
    \label{eq:lambdaIntf}
\end{multline}
where the first term again ensures the flatness condition on the axis.

In the case of the Kelvin-inverted disks, i.e. with $\nu_1 = \mathcal{K} \nu_f$ in (\ref{eq:lambdaInt1}) and (\ref{eq:lambdaInt2}), the interaction part reads
\begin{multline}
    \lambda_\text{int} =- \int_0^\infty \left[ \frac{R_+(\rho, z)}{M w - \text{i} b^2} + \frac{R_-(\rho, z)}{M w + \text{i} b^2} \right] \\
    \frac{w f(w) \dif w}{\sqrt{(b^2 + \text{i} w z)^2 - w^2\rho^2}} \,.
    \label{eq:lambdaIntfInv}
\end{multline}

\section{Particular disk solutions} \label{sec:particularSolutions}

In this section, we rederive the disk solutions from Sec.~\ref{sec:StaticThindisks} and complete their metrics by also finding the metric function $\lambda$ in some cases. Suitable disks are superposed with the Schwarzschild black hole and the respective interaction part $\lambda_\text{int}$ is computed. Finally, we also show a simpler procedure to generate the polynomial and power-law disks discussed in \cite{kotlarik2022}.

\subsection{Generalized (inverted) Morgan-Morgan disks} \label{sec:hMM}

\subsubsection{The disks}

The weight function which corresponds to the Morgan-Morgan counter-rotating family of disks reads
\begin{equation}
    f_\text{MM}^{(n)}(w \leq b) = \frac{(2n+1)!}{2^{2n} (n!)^2} \frac{\mathcal{M}}{b}  \left( 1 - \frac{w^2}{b^2} \right)^n
    \label{eq:fMM}
\end{equation}
(and zero elsewhere). Using this, we recover the potential (\ref{eq:nuMM}) by taking the real part of the integral (\ref{eq:appellconvolution}). 

New disks can easily be found by considering higher exponents in (\ref{eq:fMM}), i.e. $\frac{w^{l}}{b^{l}}$ for $l$ integers. From (\ref{eq:sigmaf}) it follows that such disks represent a certain superposition of the Morgan-Morgan disks plus a logarithmic term in density if $l$ is odd.

\begin{figure}
    \centering
    \includegraphics[width=\columnwidth]{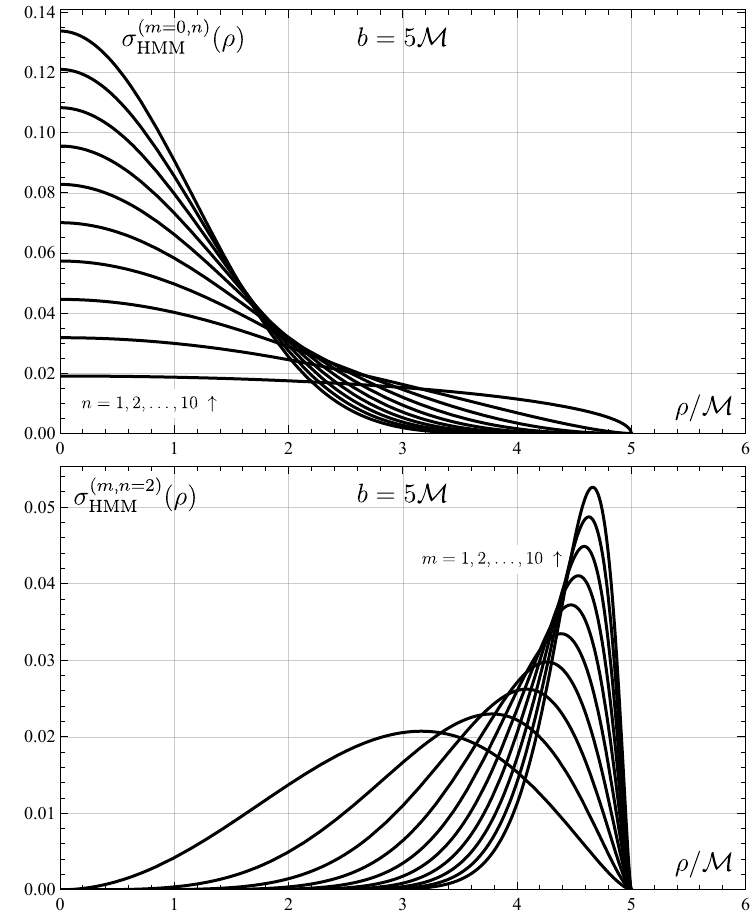}
    \caption{The first ten Newtonian-density profiles of the original Morgan-Morgan disks (\textit{top plot}) and the first ten profiles of the wider family of holey Morgan-Morgan disks (\textit{bottom plot}) with fixed $n=2$. The edge of the disk is located at $b=5\mathcal{M}$ in both plots. Increasing the parameter $n$ makes the density decrease more smoothly at the edge while increasing the parameter $m$ makes the density peak narrower and closer to the edge, finally resembling a ring rather than a disk for high values of $m$. The vertical axis is in the units of $\mathcal{M}^{-1}$.}
    \label{fig:sigmahMM}
\end{figure}

Nonetheless, there exists a more interesting and simpler generalization -- the holey Morgan-Morgan disks proposed by Letelier \cite{letelier2007}. These have densities
\begin{equation}
    \sigma_\text{hMM}^{(m,n)} (\rho \le b) = \frac{\mathcal{M}}{2 \pi b^{2+2m}} \frac{\left(\frac{3}{2}\right)_{m + n}}{m! \left(\frac{1}{2}\right)_n}\rho^{2m} \left( 1 - \frac{\rho^2}{b^2} \right)^{n - 1/2} \,,
\end{equation}
where $(a)_j = \Gamma(a + j)/\Gamma(a)$ is the Pochhammer symbol. As illustrated in Fig.~\ref{fig:sigmahMM}, contrary to the original Morgan-Morgan family, the disks with non-zero parameter $m$ have a hole in the center with the surface density exactly vanishing at $\rho = 0$. The associated weight function (\ref{eq:weightFromDensity}) reads
\begin{multline}
    f^{(m,n)}_\text{hMM}(w \leq b) = \frac{(2m)! \left(\frac{3}{2}\right)_{m + n}}{2^{2m} (m!)^2 (m + n)!} \frac{\mathcal{M}}{b} \\ 
    \sum_{j = 0}^{m + n} \binom{m+ n}{j} \frac{\left(\frac{1}{2}\right)_j}{\left(\frac{1}{2} - m\right)_j} \left(-\frac{w^2}{b^2}\right)^{j} \,.¨
    \label{eq:fhMM}
\end{multline}
The holey potential can again be obtained by integrating (\ref{eq:appellconvolution}) and taking the real part, or, by a straightforward superposition of the original Morgan-Morgan disks
\begin{equation}
    \nu_\text{hMM}^{(m,n)} = N^{(m,n)} \sum_{k=0}^{m} (-1)^k \binom{m}{k} \frac{\nu_\text{MM}^{(k + n)}}{2k  + 2n + 1} \,,
\end{equation}
with $N^{(m,n)}$ a normalization constant. The potential can be re-expressed in the basis of Legendre polynomials,
\begin{equation}
    \nu_\text{hMM}^{(m,n)} = \frac{\left(\frac{3}{2}\right)_{m + n}}{m! \left(\frac{1}{2}\right)_n} \frac{\mathcal{M}}{b} \sum_{k=0}^{m + n} C_{2k}^{(m,n)} i Q_{2k}(i \zeta) P_{2k}(\xi) \,, \label{eq:nuhMM}
\end{equation} 
where the coefficients have a slightly more complicated form
\begin{widetext}
\[   
C_{2k}^{(m,n)}  = 
    \begin{cases}
        \frac{(-1)^{k + 1}4^{-n} (4k + 1) (2n)! \left(\frac{1}{2}\right)_k }{k! (n-k)! \left(\frac{3}{2}\right)_{k+n}} \hypgeo{3}{2} \left( \genfrac{}{}{0pt}{}{ -m, \, n+ \frac{1}{2},\, n+1}{n - k + 1,\, n + k + \frac{3}{2}}; 1\right) \,, & k\leq n \\
        \frac{(-1)^{n+1} \left[\left(\frac{1}{2}\right)_k \right]^2 \left(m + n - k + 1\right)_{k-n}}{\left(\frac{1}{2}\right)_{2k} (k-n)!} \hypgeo{3}{2} \left( \genfrac{}{}{0pt}{}{k + \frac{1}{2},\, k+ 1,\, k- m-n}{2k + \frac{3}{2},\, k-n + 1}; 1\right) \,, & k > n
    \end{cases}
\]
\end{widetext}
involving the generalized hypergeometric function $\hypgeo{3}{2} \left( \genfrac{}{}{0pt}{}{ a, \,b,\,c}{d,\,e}; x\right)$ with integer and half-integer parameters evaluated at the point $x = 1$. 

For the metric function $\lambda$, we can either solve the double integration (\ref{eq:lambdaDoubleInt}), or tackle the equations for $\lambda$ directly. In the Morgan-Morgan case, the latter approach is simpler as the potential is separated in the spheroidal coordinates $\zeta$ and $\xi$. A straightforward integration of (\ref{eq:lambdaXi}) from the axis $\xi = 1$ (where $\lambda = 0$) to some general $\xi$ yields
\begin{multline}
    \lambda_\text{hMM}^{(m,n)} =\mathcal{C}_{\text{hMM}}^{(m,n)}\left[ \frac{\left(\frac{3}{2}\right)_{m + n}}{m! \left(\frac{1}{2}\right)_n} \right]^2 \frac{\mathcal{M}^2}{b^2}  (\xi^2 - 1) \left[ \mathcal{P}_{0, \text{hMM}}^{(m,n)} + \right. \\
     \left.  2\zeta \mathcal{P}_{1, \text{hMM}}^{(m,n)} \arccot(\zeta) + (\zeta^2 + 1) \mathcal{P}_{2, \text{hMM}}^{(m,n)} \arccot^{2}(\zeta) \right] \,,
     \label{eq:lambdahMM}
\end{multline}
where $\mathcal{P}_{j, \text{hMM}}^{(m,n)}$ are polynomials in $(\zeta, \xi)$, and $\mathcal{C}_{\text{hMM}}^{(m,n)}$ is a constant -- the explicit expressions are given in Appendix~\ref{app:lambdaPolynoms}. Because the Laplace equation $\Delta\nu = 0$ is also the integrability condition for $\lambda$, integrating over $\zeta$ would give the same result.

Similarly to \cite{lemos1994,semerak2003,letelier2007}, we can make the inversion (\ref{eq:kelvinOsphPot}) of the potential (\ref{eq:nuhMM}) and obtain infinite (yet finite-mass) disks stretching from the radius $\rho = b$ to infinity while leaving the central region below $b$ empty. The associated Newtonian density profiles read
\begin{equation}
    \sigma_\text{ihMM} (\rho \geq b) = \frac{b^{2m + 1}(m + n)!}{\pi^2 \left(\frac{1}{2}\right)_m \left(\frac{1}{2}\right)_n} \frac{\mathcal{M}}{\rho^{3 + 2m}} \left( 1 - \frac{b^2}{\rho^2} \right)^{n - 1/2} \,,
\end{equation}
where we again fix the normalization so that $\mathcal{M}$ stands for the total disk mass. Clearly, when $m=0$, we obtain the inverted Morgan-Morgan disks treated in \cite{lemos1994} or \cite{semerak2003}. The function $\lambda$ again follows by direct integration of the Einstein equation (\ref{eq:lambdaXi}). The result is similar to (\ref{eq:lambdahMM}) if we apply the Kelvin transformation (\ref{eq:kelvinOsphCoor}) on it, although the polynomials are different (since $\lambda$ does not transform according to (\ref{eq:kelvinOsphPot})). Specifically, the metric functions appear as
\begin{widetext}
\begin{align}
    \nu_\text{ihMM} &= \frac{2 (m + n)!}{\pi \left(\frac{1}{2}\right)_m \left(\frac{1}{2}\right)_n} \frac{\mathcal{M}}{b\sqrt{\zeta^2 + 1 - \xi^2}} \sum_{k=0}^{m + n} C_{2k}^{(m,n)} i Q_{2k}\left(\frac{i\xi}{\sqrt{\zeta^2 + 1 - \xi^2}}\right) P_{2k}\left(\frac{\zeta}{\sqrt{\zeta^2 + 1 - \xi^2}}\right) \,, \\
    \lambda_\text{ihMM}^{(m,n)} &= \mathcal{C}_{\text{ihMM}}^{(m,n)}\left[ \frac{2 (m + n)!}{\pi \left(\frac{1}{2}\right)_m \left(\frac{1}{2}\right)_n} \right]^2 \frac{\mathcal{M}^2}{b^2} \mathcal{K} \left\{(\xi^2 - 1) \left[ \mathcal{P}_{0, \text{ihMM}}^{(m,n)} + 2\zeta \mathcal{P}_{1, \text{ihMM}}^{(m,n)} \arccot(\zeta) + (\zeta^2 + 1) \mathcal{P}_{2, \text{ihMM}}^{(m,n)} \arccot^{2}(\zeta) \right] \right\} \,.
    \label{eq:lambdaihMM}
\end{align}
\end{widetext}

\begin{figure}
    \centering
    \includegraphics[width=\columnwidth]{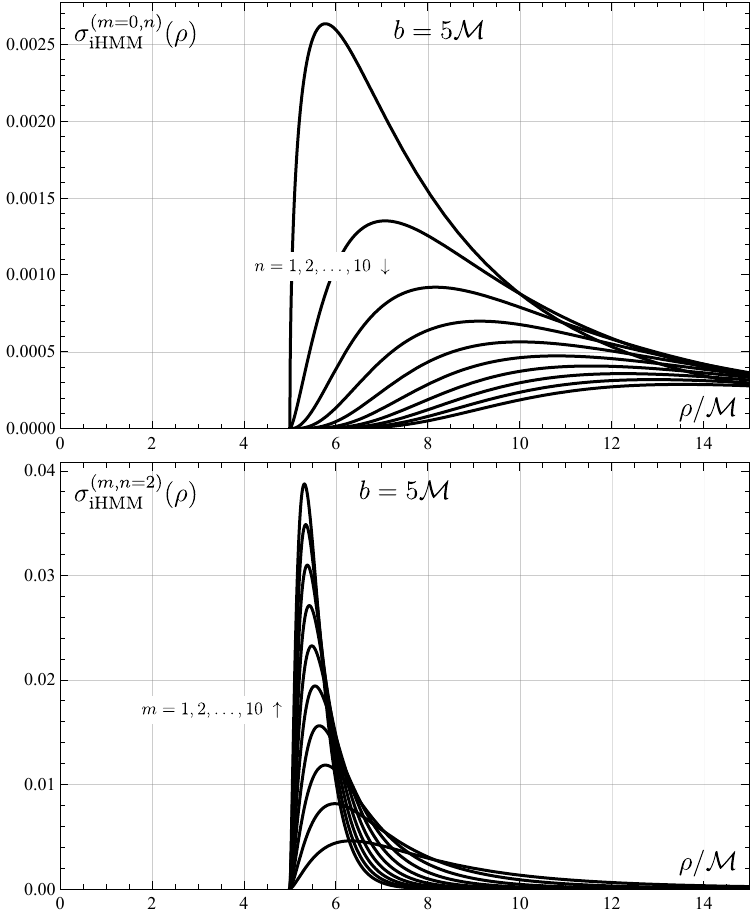}
    \caption{The first ten Newtonian-density profiles of the inverted Morgan-Morgan disks (\textit{top plot}) and the first ten profiles of the inverted holey Morgan-Morgan disks (\textit{bottom plot}) with the fixed $n=2$. As in Fig. \ref{fig:sigmahMM}, the edge of the disk is located at $b=5\mathcal{M}$ in both plots, but the disks stretch from this edge to infinity and are empty bellow $b$. Again, increasing the parameter $n$ makes the density decrease more smoothly at the edge while increasing the parameter $m$ makes the density peak narrower and closer to the edge. The vertical axis is in the units of $\mathcal{M}^{-1}$.}
    \label{fig:sigmaihMM}
\end{figure}

\subsubsection{Superposition with a black hole}

\begin{figure}
    \centering
    \includegraphics[width=\columnwidth]{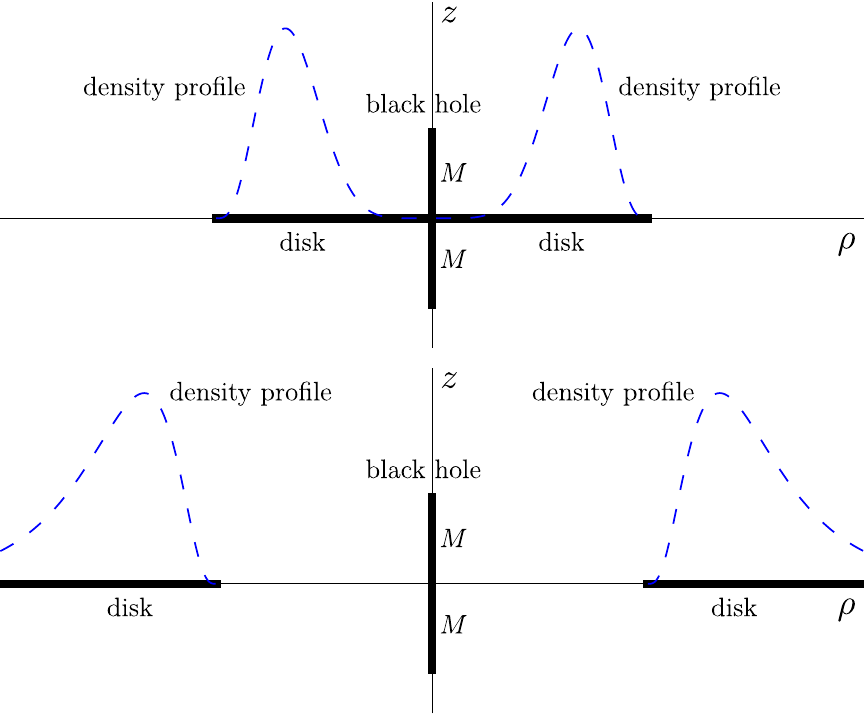}
    \caption{Schematic plot of the meridional section of the superposition of the holey Morgan-Morgan disk (\textit{top plot}) and of its inverted counterpart (\textit{bottom plot}) with a central black hole. The black hole of the mass $M$ is indicated by the thick black line placed symmetrically on the $z$ axis. The thick black lines in the equatorial plane represent the disk. The density profile is depicted by the blue dashed lines. The edge of the disk is located at $(\rho = 2.5M, z=0)$ in both plots.}
    \label{fig:hmm_sup_scheme}
\end{figure}

The holey Morgan-Morgan disks with $m \geq 1$ and their inverted versions with an arbitrary $m \geq 0$ are empty in the centre. Therefore, we can superpose them with a black hole of mass $M$ placed at the origin -- see the superposition scheme in Fig. \ref{fig:hmm_sup_scheme}. Already in Sec. \ref{subsec:solvingLambda} we discussed how to superpose sources within the Weyl class. We provided all the necessary expressions in the previous sections, yet the last piece is still missing -- the interaction part $\lambda_\text{int}$.

We will derive $\lambda_\text{int}$ from (\ref{eq:lambdaIntf}) or (\ref{eq:lambdaIntfInv}), by performing the integration over the weight function (\ref{eq:fhMM}) with the normalization adapted appropriately to the total mass $\mathcal{M}$. The general form reads
\begin{widetext}
    \begin{align}
        \lambda_\text{hMM,int}^{(m,n)} &= \mathscr{K}^{(m,n)}_{0,\text{hMM}} + \mathscr{K}^{(m,n)}_{1,\text{hMM}} \left\{ \atantwo\left[  \genfrac{}{}{0pt}{}{ b\xi - M\zeta}{ R_-(\zeta, \xi)}\right]-\atantwo\left[ \genfrac{}{}{0pt}{}{ b\xi + M\zeta}{ R_+(\zeta, \xi)} \right] \right\} + \nonumber\\
        & \qquad + \mathscr{K}^{(m,n)}_{2,\text{hMM}}\sum_\mp R_\mp(\zeta, \xi) \left[ \pm\mathscr{P}^{(m,n)}_{0,\text{hMM}}(\pm\zeta, \xi) + \mathscr{P}^{(m,n)}_{1,\text{hMM}}(\pm\zeta, \xi) \arccot(\zeta) \right] \,, \label{eq:lamndainthMM}\\
        \lambda_\text{ihMM,int}^{(m,n)} &= \mathscr{K}^{(m,n)}_{1,\text{ihMM}} \left\{ \atantwo\left[  \genfrac{}{}{0pt}{}{ b\xi - M\zeta}{ R_-(\zeta, \xi)}\right]+\atantwo\left[ \genfrac{}{}{0pt}{}{ b\xi + M\zeta}{ R_+(\zeta, \xi)} \right] \right\} + \nonumber\\
        & \qquad +  \mathscr{K}^{(m,n)}_{2,\text{ihMM}}\sum_\mp R_\mp(\zeta, \xi) \, \mathcal{K} \left\{\sqrt{1 + \zeta^2 - \xi^2}  \left[ \pm \mathscr{P}^{(m,n)}_{0,\text{ihMM}}(\pm\zeta, \xi) + \mathscr{P}^{(m,n)}_{1,\text{ihMM}}(\pm\zeta, \xi) \arccot(\zeta) \right] \right\} \,, \label{eq:lambdaintihMM}
    \end{align}
\end{widetext}
where $\atantwo(\genfrac{}{}{0pt}{}{y}{x}) \equiv \atantwo(y, x)$ denotes a 2-argument arcus tangent, $R_\pm(\zeta, \xi)$ are defined in (\ref{eq:RpRmDef}) and transformed to the oblate spheroidal coordinates (\ref{eq:CylToOsph}), $\mathscr{K}^{(m,n)}_{j, \text{(i)hMM}}$ stand for constants which depends on the disk parameters, and $\mathscr{P}^{(m,n)}_{j, \text{(i)hMM}}$ are polynomials in $\zeta$ and $\xi$. As in (\ref{eq:lambdaihMM}), we use the Kelvin transformation (\ref{eq:kelvinOsphCoor}). Explicit expressions are given in Appendix~\ref{app:lambdaPolynoms}.

\begin{figure}
    \centering
    \includegraphics[width=.7\columnwidth]{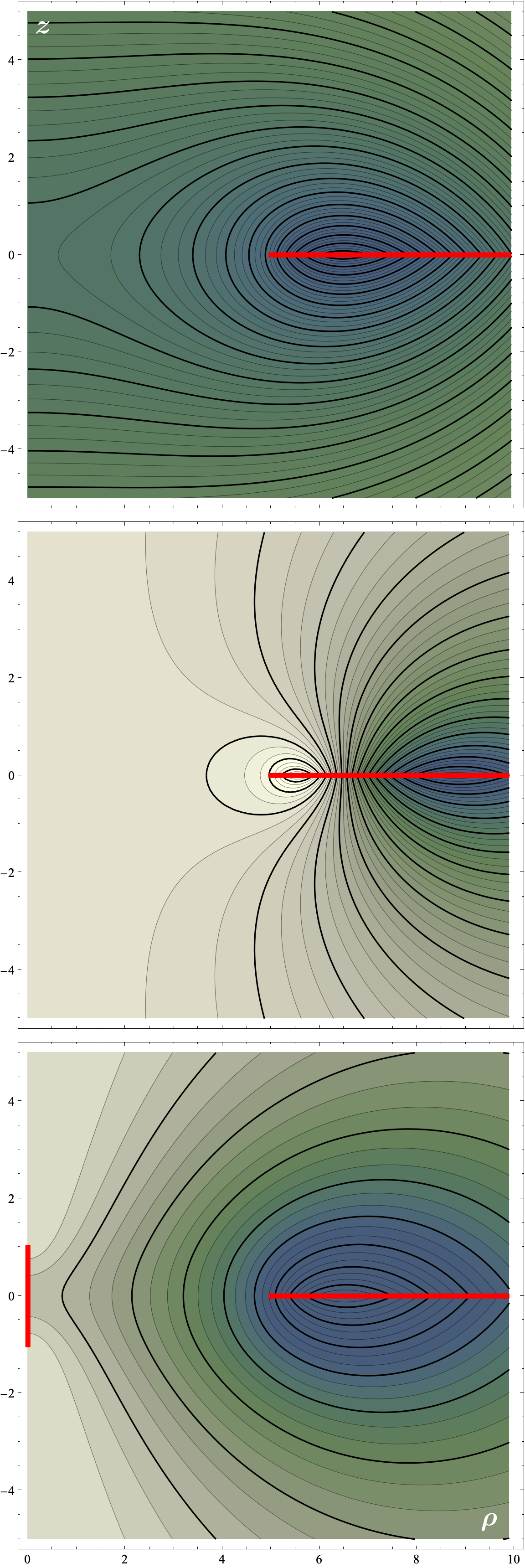}
    \caption{Meridional-section contour plots in Weyl coordinates of the potential $\nu_\text{ihMM}^{(1,2)}$ (\textit{top plot}), the second metric function $\lambda_\text{ihMM}^{(1,2)}$ (\textit{middle plot}), and the interaction part $\lambda_\text{int,ihMM}^{(1,2)}$ (\textit{bottom plot}) of the inverted holey Morgan-Morgan disk with $(m = 1, n = 2)$ superposed with the Schwarzschild black hole of mass $M$. The disk's inner edge is located at $(\rho = 5M, z = 0)$ and its mass is $\mathcal{M} = 2M$. The disk is indicated by the thick horizontal red line, while the black hole is indicated by the thick vertical red line. In the middle plot, the largest negative values are coloured deep blue; they increase through green to light brown (indicating zero value) up to positive values indicated by white. The values range from about $-0.32$ to $-0.17$ in the top plot, from about $-0.030$ to $0.0066$ in the middle plot, and from about $-0.044$ to 0 in the bottom plot. Both axes are in the units of~$M$.}
\end{figure}

\subsection{Kuzmin-Toomre \& Vogt-Letelier disks}

The suggested method is general in the sense that most of the known disk solutions can be reformulated in terms of the weight function and integration (\ref{eq:appellconvolution}). From (\ref{eq:sigmaf}), we easily obtain the weight function associated with the Kuzmin-Toomre disks (\ref{eq:sigmaKT}),
\begin{equation}
    f_\text{KT}^{(n)} (w) = \frac{2^{2 n + 1} (n!)^2}{\pi (2n)!} \frac{\mathcal{M} b^{2n + 1}}{(w^2 + b^2)^{1 + n}} \,, \qquad n \in \mathbb{N}_0 \,.
\end{equation}
Note that these disks are infinite, thus $w$ ranges through the full interval $\left[ 0, \infty \right)$, with $b$ just a length parameter not directly indicating the location of the disk edge. A simple generalization to half-integer exponents is possible as well,
\begin{equation}
    f_\text{gKT}^{(n)} (w) = \frac{n \, b^{2n}(2n)!}{2^{2n - 1}(n!)^2} \frac{\mathcal{M}}{(w^2 + b^2)^{n + 1/2}} \,,
\end{equation}
resulting in Newtonian surface densities
\begin{equation}
    \sigma_\text{gKT}^{(n)} (\rho) = \frac{n \, b^{2n}}{\pi}  \frac{\mathcal{M}}{(\rho^2 + b^2)^{1 + n}}  \,.
\end{equation}
However, the potential (\ref{eq:appellconvolution}) is then given in terms of complete elliptic integrals, which we do not list here. 

The weight function of the Vogt-Letelier disk family~(\ref{eq:VLdensity}),
\begin{multline}
    f_\text{VL}^{(m,n)}(w) = \frac{2}{\pi (2 m + 1)} \frac{m!}{n!} \frac{\left( \frac{3}{2} \right)_{m + n}}{\left[ \left( \frac{1}{2} \right)_m \right]^2} \frac{\mathcal{M}}{b} \left(\frac{b}{w} \right)^{2m + 2} \times \\
    \times \hypgeo{2}{1} \left(\genfrac{}{}{0pt}{}{1 + m, \, \frac{3 + 2m + 2n}{2}}{\frac{3 + 2m}{2}} ; -\frac{b^2}{w^2} \right) \,,
\end{multline}
can be reduced to
\begin{equation}
    f_\text{VL}^{(m,n)}(w) \propto \frac{\mathscr{P}_\text{VL}^{(m,n)}(w)}{(w^2 + b^2)^{m + n + 1}} \,,
    \label{eq:VLweightfunctionPropto}
\end{equation}
where $\mathscr{P}_\text{VL}^{(m,n)}(w)$ is an even polynomial of the order $2n$ in~$w$. Again, one can also consider half-integer exponents of $(w^2 + b^2)^{-m - n - 1/2}$ in (\ref{eq:VLweightfunctionPropto}) and compute the respective potential from (\ref{eq:appellconvolution}). The result corresponds to the Newtonian surface densities (\ref{eq:VLdensity}), wherein the denominator there appears $(\rho^2 + b^2)^{m + n + 1}$ rather than a half-integer exponent. It again involves complete elliptic integrals.

In order to find the second metric function $\lambda$, one may use the approach presented above, including the interaction part $\lambda_\text{int}$ necessary in superpositions with a black hole, yet it is easier to follow the direct approach like in \cite{bicak1993,kotlarik2022a}.

\subsection{Polynomial and power-law disks}

\subsubsection{The weight function}

The polynomial, power-law and bump disks derived in \cite{kotlarik2022} can also be recast in the present formalism. In particular, the elementary density terms $\sigma^{(2l)}_\text{elem}(\rho \leq b) = \rho^{2l}$, $l \in \mathbb{N}_0$, translate to the weight function 
\begin{align}
    f_\text{elem}^{(2l)} (w \leq b) &= - 2 w^{2l + 1} B _{\frac{w^2}{b^2}} \left( -\frac{1}{2} - l,\, \frac{1}{2} \right)\\
    &= \frac{b^{2l}\sqrt{b^2 - w^2}}{\pi^2(2l + 1)} \sum_{k = 0}^{l} \frac{(-l)_k}{\left( \frac{1}{2} -l \right)_k} \left( \frac{w}{b} \right)^{2k} \,,
\end{align}
where $B_x(c,d)$ is the incomplete Beta function. By evaluating the integral (\ref{eq:appellconvolution}) and taking the real part, we reproduce the results of \cite{kotlarik2022} obtained by direct integration of the axially symmetric Green function (\ref{eq:conwayIntegral}). Although the computation of $\lambda$ is still beyond our reach, for the power-law disks superposed with a black hole the computation of the interaction part $\lambda_\text{int}$ is feasible actually.

\subsubsection{Simpler recurrence relation for the potentials in \cite{kotlarik2022}}

Before finishing this section, let us add a more straightforward way how to derive the potential-density pairs in~\cite{kotlarik2022}. In that paper, we used the approach proposed by~\cite{conway2000} and explicitly integrated the axially symmetric Green function (\ref{eq:conwayIntegral}). Considering the elementary density terms $\sigma(\rho) = \rho^{2l}$, the problem reduces to the computation of the Bessel-Laplace integrals
\begin{equation}
    \mathcal{I}_{(\alpha, \beta, \gamma)} = \int_0^\infty s^\alpha J_\beta(s b) J_\gamma     (s \rho) e^{-s |z|} \dif s \;.
\end{equation}
Various recurrence relations can be used to explicitly calculate $\mathcal{I}_{(\alpha, \beta, \gamma)}$. In what follows, we show that it is not necessary to use the lengthy recurrence relations from \cite{kotlarik2022}. Actually, we have found a simpler recipe inferred from relations between solutions to the generalized $2s + 1$ dimensional Laplace (Poisson) equations.

The key observation is that if $\nu$ is an axially symmetric solution to the Laplace equation, then $\od[n]{\nu}{z}$ for $n \in \mathbb{N}$ is also a solution. Let us consider the equation for the generalized axially symmetric Laplace (Poisson) equation in a flat space of the dimension $(2s+3)$ (for integer and half-integer spin values $s\geq -1/2$),
\begin{equation}
    \pd[2]{\nu_{(s)}}{z} + \pd[2]{\nu_{(s)}}{\rho} + \frac{1 + 2s}{\rho} \pd{\nu_{(s)}}{\rho} \equiv \Delta_s \nu_{(s)}  = \sigma(\rho, z) \,.
\label{eq:GASP-Laplace}
\end{equation}
Now, the generalized Laplace equation can be also extended to spins $s<-1/2$, since it holds
\begin{equation}
    \Delta_{-s} \nu_{(-s)} = \rho^{2s}\,\Delta_s \left[\rho^{-2s}\nu_{(-s)} \right]\,.
\end{equation}
The Kelvin transformation can be extended to work for arbitrary $s$ as well, because
\begin{equation}
    \mathcal{K} \nu_{(s)}(\rho, z) \equiv \left(\frac{b}{\rho^2 + z^2}\right)^{2s + 1} \nu_{(s)} \left( \frac{b^2 \rho}{\rho^2 + z^2}, \frac{b^2 z}{\rho^2 + z^2} \right)
\end{equation}
solves (\ref{eq:GASP-Laplace}) if $\nu_{(s)}$ is a solution (with the density adjusted appropriately). 

One can find identities between the different-$s$ solutions of (\ref{eq:GASP-Laplace}). If $\nu_{(s)}$ satisfies
\begin{equation}
    \Delta_s \nu_{(s)} = \sigma(\rho, z) \,,
\end{equation}
then it holds
\begin{equation} \label{eq:GASP-idUp}
    \Delta_{s+1} \left[ \frac{1}{\rho} \dpd{\nu_{(s)}}{\rho} \right] = \frac{1}{\rho} \dpd{}{\rho} \Delta_s \nu_{(s)} = \frac{1}{\rho} \dpd{\sigma(\rho,z)}{\rho}
\end{equation}
and
\begin{align}
    \Delta_{s - 1} \left[ \frac{1}{\rho^{2s - 1}} \dpd{\rho^{2s} \nu_{(s)}}{\rho} \right] &= \frac{1}{\rho^{2s - 1}} \dpd{}{\rho}\rho^{2s} \Delta_s \nu_{(s)} \nonumber\\
    &= \frac{1}{\rho^{2s - 1}} \dpd{}{\rho} \rho^{2s} \sigma(\rho, z) \,.  \label{eq:GASP-idDown}
\end{align}
Therefore, from known solutions $\nu_{(s)}$ of the Laplace equation, we can generate new solutions $\nu_{(s \pm 1)}$ with the source density distributions given by (\ref{eq:GASP-idUp}) and (\ref{eq:GASP-idDown}). 

It is then natural to introduce spin-raising and spin-lowering operators as follows,
\begin{equation} \label{eq:GASP-spinOperators}
    \mathcal{S}_{s+1}^s \nu_{(s)} \equiv \frac{\kappa^2}{\rho} \dpd{\nu_{(s)}}{\rho} \quad \text{and} \quad \mathcal{S}_{s-1}^s \nu_{(s)} \equiv \frac{1}{\rho^{2s - 1}} \dpd{\rho^{2s} \nu_{(s)}}{\rho} \,,
\end{equation}
where $\kappa$ is a constant of the dimension of length. As a seed potential-density pair we choose one of the solutions provided in \cite{kotlarik2022}, in particular the one which describes the field of a thin infinite annular disk extending from $\rho = b$ to $\rho = \infty$ lying in the $z = 0$ plane. Namely\footnote{Note that normalization is different from \cite{kotlarik2022}.}
\begin{widetext}
    \begin{align}
        \nu_{(0)}^{(-3)}(\rho, z) =& \frac{\sigma_c \, b^2}{2\pi} \left[\frac{2\pi b |z|}{(\rho^2 + z^2)^{3/2}} H\left(b- \frac{\rho \, b^2}{\rho^2 + z^2}\right) - \frac{4 \sqrt{\rho b}}{k(\rho^2 + z^2)} E(k) - \frac{(\rho^2 + \rho b + z^2)(\rho^2 - \rho b + z^2)k}{\sqrt{\rho b} \, (\rho^2 + z^2)^2} K(k) - \right. \nonumber \\
        &\left. - \frac{\rho^2 - \rho b + z^2}{\rho^2 + \rho b + z^2} \frac{b^2 \, z^2 \, k}{(\rho^2 + z^2)^2 \, \sqrt{\rho b}} \Pi \left( \frac{4 b \rho (\rho^2 + z^2)}{(\rho^2 + b \rho + z^2)^2}, k \right) \right] \\
        \sigma_{(0)}^{(-3)}(\rho, z) &= \frac{\sigma_c}{2\pi}\frac{b^3}{\rho^3} H(\rho - b) \delta(z) \,,
    \end{align}
\end{widetext}
where $\sigma_c$ is a constant with the dimension of surface density (inverse length), $H$ is the Heaviside step function, and $K, E, \Pi$ are the complete elliptic integrals of the first, the second, and the third kind with the modulus
\begin{equation}
    k = \frac{2 \sqrt{\rho b}}{\sqrt{(\rho + b)^2 + z^2}} \,.
\end{equation}
Applying the spin operators (\ref{eq:GASP-spinOperators}) to the potentials of the type $\nu_{(0)}^{(-3)}$ introduces derivatives of the Heaviside step function, which adds an artificial distributional source to the edge of the disk. To get rid of that, one has to subtract the potential of a ring with a uniform density, satisfying
\begin{equation}
    \Delta_s \nu_{(s)}^{(\delta)} = \frac{1}{2\pi} \delta(\rho - b) \delta(z) \,.
    \label{eq:laplacesRing}
\end{equation}
The recurrence relations are thus obtained
\begin{align}
    \nu_{(0)}^{(- 3 - 2l)} &= -\frac{1}{2l + 1} \left[ \mathcal{S}^{-1}_0 \nu_{(-1)}^{(- 1 - 2l)} - \left(\frac{\kappa}{b}\right)^{2l + 2}  \nu_{(0)}^{(\delta)} \right] \,,\\
    \nu_{(-1)}^{(-2l - 1)} &= -\frac{1}{2l + 1} \left[ \mathcal{S}_{-1}^{0} \nu_{(0)}^{(-2l - 1)} -  \left(\frac{\kappa}{b}\right)^{2l + 2} \nu_{(-1)}^{(\delta)}  \right] \,,
\end{align}
where $\kappa^2 = b^3\sigma_c$, and $\nu_{(0)}^{(\delta)}, \nu_{(-1)}^{(\delta)}$ are solutions of (\ref{eq:laplacesRing}) with $s = 0,-1$, respectively, namely
\begin{align}
    \nu_{(0)}^{(\delta)} &= - \frac{b}{\pi} \frac{K(k)}{\sqrt{(\rho - b)^2 + z^2}} \,.\\
    \nu_{(-1)}^{(\delta)} &= \frac{1}{2\pi b} \left[ \sqrt{(\rho - b)^2 + z^2} E(k) - \right. \\
    & \qquad \qquad \qquad \qquad\left. - \frac{(\rho^2 + b^2 + z^2)K(k)}{\sqrt{(\rho - b)^2 + z^2}} \right] \,.
\end{align}

In this way, we design disks with the surface density profile proportional to $\rho^{-2l - 3}$ from the seed solution $\nu_{(0)}^{(-3)}$. The Kelvin transformation then automatically provides $\nu_{(0)}^{(2l)}$, thus the densities proportional to $\rho^{2l}$. This construction works for any monotonous density profile, but a closed-form formula for the density $\rho^{-2}$ is not yet known.

\section{Conclusions}
\label{sec:conclusions}

We have revisited the topic of thin disks as sources of the Weyl class of spacetimes and proposed a new general method for obtaining their density and potential. The method can be understood in two ways: (\textit{i}) Kuzmin's idea \cite{kuzmin1956} and the Appell's trick \cite{appell1887} are combined in such way that the gravitational field is obtained from the field of a line distribution of matter described by the weight function $f(w)$ (an Abel transform of the Newtonian surface density) placed on the imaginary extension of the $z$ axis, cut and mirrored with respect to the equatorial plane; (\textit{ii}) a certain superposition (convolution) of Appell rings of radius $w$ is made as weighted by the weight function $f(w)$. We showed on particular examples that the procedure is capable of reproducing various thin-disk solutions known from the literature. Moreover, it has proved useful in deriving the second metric function $\lambda$ which is only rarely known explicitly. Indeed, we provided the whole metric in closed forms for the general holey Morgan-Morgan disks and their superposition with a Schwarzschild black hole. Finally, we also showed that the polynomial and power-law disks treated in \cite{kotlarik2022} can be obtained by an easier procedure.

To conclude, let us acknowledge that Letelier \& Oliveira~\cite{letelier1987} actually discovered the relation (\ref{eq:appellconvolution}) for the zeroth member of the Morgan-Morgan family. Also, Klein~\cite{klein1997} used a similar type of integration to construct inverted isochrone disks around Schwarzschild black holes, although he did not give the explicit form of the potential nor the second metric function $\lambda$. Anyway, the method presented here is completely general and can be applied to any thin-disk solution within the Weyl class of spacetimes.

\begin{acknowledgments}
We are thankful for support from the grant GACR 21-11268S of the Czech Science Foundation.
\end{acknowledgments}

\onecolumngrid

\appendix

\section{The second metric function $\lambda$ for the (inverted) generalized Morgan-Morgan disks}
\label{app:lambdaPolynoms}

In the following, we list explicit results for the second metric function $\lambda_\text{disk}$ for the (inverted) holey Morgan-Morgan disks, and the interaction part $\lambda_\text{int}$ for their superposition with the black hole. We also provide a Mathematica notebook containing full expressions in the Supplement Material.

\subsection{Holey Morgan-Morgan disks}

\paragraph{Qudratic $\lambda_\text{disk}$:} A general expression for the second metric function of the holey Morgan-Morgan disks (\ref{eq:lambdahMM}) reads
\begin{equation}
    \lambda_\text{hMM}^{(m,n)} =\mathcal{C}_{\text{hMM}}^{(m,n)}\left[ \frac{\left(\frac{3}{2}\right)_{m + n}}{m! \left(\frac{1}{2}\right)_n} \right]^2 \frac{\mathcal{M}^2}{b^2}  (\xi^2 - 1) \left[ \mathcal{P}_{0, \text{hMM}}^{(m,n)} + 2\zeta \mathcal{P}_{1, \text{hMM}}^{(m,n)} \arccot(\zeta) + (\zeta^2 + 1) \mathcal{P}_{2, \text{hMM}}^{(m,n)} \arccot^{2}(\zeta) \right] \,, \label{eq:app:lambdahMM}
\end{equation}
where $\mathcal{C}_{\text{hMM}}^{(m,n)}$ are numerical constants and $\mathcal{P}_{j, \text{hMM}}^{(m,n)}$ are polynomials in $(\zeta, \xi)$. The first few members read

\begingroup
\allowdisplaybreaks

\begin{align}
    \mathcal{C}_{\text{hMM}}^{(0,1)} &= \frac{1}{16} \\
    \mathcal{P}_{0, \text{hMM}}^{(0,1)} &= (9 \zeta ^2+4) \xi ^2-\zeta ^2+4\\
    \mathcal{P}_{1, \text{hMM}}^{(0,1)} &= -(9 \zeta ^2+7) \xi ^2+\zeta ^2-1\\
    \mathcal{P}_{2, \text{hMM}}^{(0,1)} &= \zeta ^2 (9 \xi ^2-1)+\xi ^2-1\\
    \mathcal{C}_{\text{hMM}}^{(0,2)} &= \frac{1}{2048} \\
    \mathcal{P}_{0,\text{hMM}}^{(0,2)} &=11025\zeta^6\xi^6-11475\zeta^6\xi^4+2835\zeta^6\xi^2-81\zeta^6+15150\zeta^4\xi^6-10890\zeta^4\xi^4+1098\zeta^4\xi^2+18\zeta^4+4945\zeta^2\xi^6 \nonumber\\
    &\quad -723\zeta^2\xi^4-45\zeta^2\xi^2-81\zeta^2+256\xi^6+256\xi^4+256\xi^2+256\\
    \mathcal{P}_{1,\text{hMM}}^{(0,2)}&=-3(3675\zeta^6\xi^6-3825\zeta^6\xi^4+945\zeta^6\xi^2-27\zeta^6+6275\zeta^4\xi^6-4905\zeta^4\xi^4+681\zeta^4\xi^2-3\zeta^4+3005\zeta^2\xi^6-1111\zeta^2\xi^4 \nonumber\\
    &\quad -105\zeta^2\xi^2+3\zeta^2+357\xi^6+113\xi^4+15\xi^2+27)\\
    \mathcal{P}_{2,\text{hMM}}^{(0,2)} &=9(1225\zeta^6\xi^6-1275\zeta^6\xi^4+315\zeta^6\xi^2-9\zeta^6+1275\zeta^4\xi^6-785\zeta^4\xi^4+17\zeta^4\xi^2+5\zeta^4+315\zeta^2\xi^6-17\zeta^2\xi^4-47\zeta^2\xi^2 \nonumber\\
    &\quad +5\zeta^2+9\xi^6+5\xi^4-5\xi^2-9) \\
    \mathcal{C}_{\text{hMM}}^{(1,1)} &= \frac{1}{6144}\\
    \mathcal{P}_{0,\text{hMM}}^{(1,1)}&=33075\zeta^6\xi^6-34425\zeta^6\xi^4+8505\zeta^6\xi^2-243\zeta^6+45450\zeta^4\xi^6-52830\zeta^4\xi^4+14814\zeta^4\xi^2-522\zeta^4+14835\zeta^2\xi^6 \nonumber\\
    &\quad -20409\zeta^2\xi^4+5625\zeta^2\xi^2-51\zeta^2+768\xi^6-1280\xi^4+256\xi^2+256 \\
    \mathcal{P}_{1,\text{hMM}}^{(1,1)}&=-3(11025\zeta^6\xi^6-11475\zeta^6\xi^4+2835\zeta^6\xi^2-81\zeta^6+18825\zeta^4\xi^6-21435\zeta^4\xi^4+5883\zeta^4\xi^2-201\zeta^4+9015\zeta^2\xi^6 \nonumber\\
    &\quad -11653\zeta^2\xi^4+3525\zeta^2\xi^2-119\zeta^2+1071\xi^6-1645\xi^4+557\xi^2+17) \\
    \mathcal{P}_{2,\text{hMM}}^{(1,1)}&=3(11025\zeta^6\xi^6-11475\zeta^6\xi^4+2835\zeta^6\xi^2-81\zeta^6+11475\zeta^4\xi^6-13785\zeta^4\xi^4+3993\zeta^4\xi^2-147\zeta^4+2835\zeta^2\xi^6 \nonumber\\
    &\quad -3993\zeta^2\xi^4+1497\zeta^2\xi^2-83\zeta^2+81\xi^6-147\xi^4+83\xi^2-17) \\
    \mathcal{C}_{\text{hMM}}^{(1,2)} &= \frac{1}{983040} \\
    \mathcal{P}_{0,\text{hMM}}^{(1,2)} &= 60031125\zeta^{10}\xi^{10}-121550625\zeta^{10}\xi^8+84341250\zeta^{10}\xi^6-23231250\zeta^{10}\xi^4+2165625\zeta^{10}\xi^2-28125\zeta^{10} \nonumber\\
    &\quad +141561000\zeta^8\xi^{10}-278359200\zeta^8\xi^8+186202800\zeta^8\xi^6-48904800\zeta^8\xi^4+4276200\zeta^8\xi^2-52800\zeta^8 \nonumber\\
    &\quad +115519950\zeta^6\xi^{10}-218461950\zeta^6\xi^8+137270700\zeta^6\xi^6-32294300\zeta^6\xi^4+2275750\zeta^6\xi^2-17030\zeta^6 \nonumber\\
    &\quad +37963800\zeta^4\xi^{10}-68050080\zeta^4\xi^8+37925520\zeta^4\xi^6-6760800\zeta^4\xi^4+220440\zeta^4\xi^2-960\zeta^4 \nonumber\\
    &\quad +4282845\zeta^2\xi^{10}-7093665\zeta^2\xi^8+2940690\zeta^2\xi^6-127890\zeta^2\xi^4+225\zeta^2\xi^2-2205\zeta^2 \nonumber\\
    &\quad +81920\xi^{10}-114688\xi^8+8192\xi^6+8192\xi^4+8192\xi^2+8192 \\
    \mathcal{P}_{1,\text{hMM}}^{(1,2)}&=-15(4002075\zeta^{10}\xi^{10}-8103375\zeta^{10}\xi^8+5622750\zeta^{10}\xi^6-1548750\zeta^{10}\xi^4+144375\zeta^{10}\xi^2-1875\zeta^{10} \nonumber \\
    &\quad +10771425\zeta^8\xi^{10}-21258405\zeta^8\xi^8+14287770\zeta^8\xi^6-3776570\zeta^8\xi^4+333205\zeta^8\xi^2-4145\zeta^8+10491390\zeta^6\xi^{10} \nonumber \\
    &\quad -20029590\zeta^6\xi^8+12789420\zeta^6\xi^6-3102060\zeta^6\xi^4+233910\zeta^6\xi^2-2142\zeta^6+4445490\zeta^4\xi^{10}-8119146\zeta^4\xi^8 \nonumber \\
    &\quad +4737204\zeta^4\xi^6-941460\zeta^4\xi^4+40506\zeta^4\xi^2+222\zeta^4+763175\zeta^2\xi^{10}-1309915\zeta^2\xi^8+635126\zeta^2\xi^6-63046\zeta^2\xi^4 \nonumber \\
    &\quad -5485\zeta^2\xi^2+113\zeta^2+36525\xi^{10}-57009\xi^8+16434\xi^6+3918\xi^4-15\xi^2+147) \\
    \mathcal{P}_{2,\text{hMM}}^{(1,2)} &= 45(1334025\zeta^{10}\xi^{10}-2701125\zeta^{10}\xi^8+1874250\zeta^{10}\xi^6-516250\zeta^{10}\xi^4+48125\zeta^{10}\xi^2-625\zeta^{10}+2701125\zeta^8\xi^{10} \nonumber \\
    &\quad -5285385\zeta^8\xi^8+3513090\zeta^8\xi^6-914690\zeta^8\xi^4+78985\zeta^8\xi^2-965\zeta^8+1874250\zeta^6\xi^{10}-3513090\zeta^6\xi^8 \nonumber \\
    &\quad +2170980\zeta^6\xi^6-493060\zeta^6\xi^4+31730\zeta^6\xi^2-154\zeta^6+516250\zeta^4\xi^{10}-914690\zeta^4\xi^8+493060\zeta^4\xi^6-74500\zeta^4\xi^4 \nonumber \\
    &\quad -1918\zeta^4\xi^2+230\zeta^4+48125\zeta^2\xi^{10}-78985\zeta^2\xi^8+31730\zeta^2\xi^6+1918\zeta^2\xi^4-1247\zeta^2\xi^2-5\zeta^2+625\xi^{10}-965\xi^8 \nonumber \\
    &\quad +154\xi^6+230\xi^4+5\xi^2-49) \\
    \mathcal{C}_{\text{hMM}}^{(2,1)} &= \frac{1}{983040} \\
    \mathcal{P}_{0,\text{hMM}}^{(2,1)} &= 60031125 \zeta ^{10} \xi ^{10}-121550625 \zeta ^{10} \xi ^8+84341250 \zeta ^{10} \xi ^6-23231250 \zeta ^{10}\xi ^4+2165625 \zeta ^{10} \xi ^2-28125 \zeta ^{10} \nonumber \\
    &\quad +141561000 \zeta ^8 \xi ^{10}-313286400 \zeta ^8 \xi^8+239727600 \zeta ^8 \xi ^6-73701600 \zeta ^8 \xi ^4+7775400 \zeta ^8 \xi ^2-117600 \zeta ^8 \nonumber \\
    &\quad +115519950\zeta ^6 \xi ^{10}-283629150 \zeta ^6 \xi ^8+242657100 \zeta ^6 \xi ^6-84343100 \zeta ^6 \xi ^4+10217350\zeta ^6 \xi ^2-184070 \zeta ^6 \nonumber \\
    &\quad +37963800 \zeta ^4 \xi ^{10}-105255360 \zeta ^4 \xi ^8+102564240 \zeta ^4\xi ^6-40778400 \zeta ^4 \xi ^4+5627160 \zeta ^4 \xi ^2-113760 \zeta ^4 \nonumber \\
    &\quad +4282845 \zeta ^2 \xi^{10}-13747905 \zeta ^2 \xi ^8+15665970 \zeta ^2 \xi ^6-7221810 \zeta ^2 \xi ^4+1024065 \zeta ^2 \xi^2-3165 \zeta ^2+81920 \xi ^{10} \nonumber \\
    &\quad -311296 \xi ^8+425984 \xi ^6-229376 \xi ^4+16384 \xi ^2+16384 \\
    \mathcal{P}_{1,\text{hMM}}^{(2,1)}&=-15(4002075\zeta^{10}\xi^{10}-8103375\zeta^{10}\xi^8+5622750\zeta^{10}\xi^6-1548750\zeta^{10}\xi^4+144375\zeta^{10}\xi^2-1875\zeta^{10} \nonumber \\
    &\quad +10771425\zeta^8\xi^{10}-23586885\zeta^8\xi^8+17856090\zeta^8\xi^6-5429690\zeta^8\xi^4+566485\zeta^8\xi^2-8465\zeta^8 \nonumber \\
    &\quad +10491390\zeta^6\xi^{10}-25150230\zeta^6\xi^8+21004620\zeta^6\xi^6-7123020\zeta^6\xi^4+841110\zeta^6\xi^2-14718\zeta^6+4445490\zeta^4\xi^{10} \nonumber \\
    &\quad -11840682\zeta^4\xi^8+11071188\zeta^4\xi^6-4240500\zeta^4\xi^4+571034\zeta^4\xi^2-11650\zeta^4+763175\zeta^2\xi^{10}-2302555\zeta^2\xi^8 \nonumber \\
    &\quad +2459990\zeta^2\xi^6-1083494\zeta^2\xi^4+165811\zeta^2\xi^2-3439\zeta^2+36525\xi^{10}-127953\xi^8+160850\xi^6-83794\xi^4 \nonumber \\
    &\quad +14161\xi^2+211)\\ 
    \mathcal{P}_{2,\text{hMM}}^{(2,1)} &= 15(4002075\zeta^{10}\xi^{10}-8103375\zeta^{10}\xi^8+5622750\zeta^{10}\xi^6-1548750\zeta^{10}\xi^4+144375\zeta^{10}\xi^2-1875\zeta^{10}+8103375\zeta^8\xi^{10} \nonumber \\
    &\quad -18184635\zeta^8\xi^8+14107590\zeta^8\xi^6-4397190\zeta^8\xi^4+470235\zeta^8\xi^2-7215\zeta^8+5622750\zeta^6\xi^{10}-14107590\zeta^6\xi^8 \nonumber \\
    &\quad +12349260\zeta^6\xi^6-4398060\zeta^6\xi^4+546870\zeta^6\xi^2-10158\zeta^6+1548750\zeta^4\xi^{10}-4397190\zeta^4\xi^8+4398060\zeta^4\xi^6 \nonumber \\
    &\quad -1818540\zeta^4\xi^4+269094\zeta^4\xi^2-6318\zeta^4+144375\zeta^2\xi^{10}-470235\zeta^2\xi^8+546870\zeta^2\xi^6-269094\zeta^2\xi^4+50307\zeta^2\xi^2 \nonumber \\
    &\quad -1711\zeta^2+1875\xi^{10}-7215\xi^8+10158\xi^6-6318\xi^4+1711\xi^2-211) \\
    \mathcal{C}_{\text{hMM}}^{(2,2)} &= \frac{1}{21139292160} \\
    \mathcal{P}_{0,\text{hMM}}^{(2,2)} &= 15978784696875 \zeta^{14} \xi^{14}-48220421374125 \zeta^{14} \xi^{12}+56556703578375 \zeta^{14} \xi^{10}-32478448550625 \zeta^{14} \xi^8 \nonumber \\
    &\quad +9431153510625 \zeta^{14} \xi^6-1280535834375 \zeta^{14} \xi^4+63814078125 \zeta^{14} \xi^2-472696875 \zeta^{14}+53546682939750 \zeta^{12} \xi^{14} \nonumber \\
    &\quad -163921910402250 \zeta^{12} \xi^{12}+195401453766750 \zeta^{12} \xi^{10}-114310631531250 \zeta^{12} \xi^8+33912309611250 \zeta^{12} \xi^6 \nonumber \\
    &\quad -4721464518750 \zeta^{12} \xi^4+242354306250 \zeta^{12} \xi^2-1878843750 \zeta^{12}+70144588639125 \zeta^{10} \xi^{14} \nonumber \\
    &\quad -218331077008275 \zeta^{10} \xi^{12}+264573915831225 \zeta^{10} \xi^{10}-157245484449375 \zeta^{10} \xi^8+47328582215775 \zeta^{10} \xi^6 \nonumber \\
    &\quad -6667480157625 \zeta^{10} \xi^4+344764777875 \zeta^{10} \xi^2-2706591125 \zeta^{10}+45300791358900 \zeta^8 \xi^{14} \nonumber \\
    &\quad -143774916446700 \zeta^8 \xi^{12}+177053751144900 \zeta^8 \xi^{10}-106350197324700 \zeta^8 \xi^8+32049674367900 \zeta^8 \xi^6 \nonumber \\
    &\quad -4445363304900 \zeta^8 \xi^4+219341509100 \zeta^8 \xi^2-1563454900 \zeta^8+14914843730325 \zeta^6 \xi^{14}-48441884967315 \zeta^6 \xi^{12} \nonumber \\
    &\quad +60540898489785 \zeta^6 \xi^{10}-36386505134175 \zeta^6 \xi^8+10693987893855 \zeta^6 \xi^6-1374141802425 \zeta^6 \xi^4 \nonumber \\
    &\quad +55823145875 \zeta^6 \xi^2-227556245 \zeta^6+2318953337190 \zeta^4 \xi^{14}-7743583874250 \zeta^4 \xi^{12}+9790161391710 \zeta^4 \xi^{10} \nonumber \\
    &\quad -5770162221810 \zeta^4 \xi^8+1558297174770 \zeta^4 \xi^6-156345952350 \zeta^4 \xi^4+2730718410 \zeta^4 \xi^2-9285990 \zeta^4 \nonumber \\
    &\quad +133728684075 \zeta^2 \xi^{14}-462058266285 \zeta^2 \xi^{12}+586633472775 \zeta^2 \xi^{10}-323550305505 \zeta^2 \xi^8+66781739745 \zeta^2 \xi^6 \nonumber \\
    &\quad -1531728135 \zeta^2 \xi^4+6065325 \zeta^2 \xi^2-9661995 \zeta^2+1321205760 \xi^{14}-4718592000 \xi^{12}+5851054080 \xi^{10} \nonumber \\
    &\quad -2604662784 \xi^8+37748736 \xi^6+37748736 \xi^4+37748736 \xi^2+37748736 \\
    \mathcal{P}_{1,\text{hMM}}^{(2,2)}&=-105(152178901875\zeta^{14}\xi^{14}-459242108325\zeta^{14}\xi^{12}+538635272175\zeta^{14}\xi^{10}-309318557625\zeta^{14}\xi^8 \nonumber \\
    &\quad +89820509625\zeta^{14}\xi^6-12195579375\zeta^{14}\xi^4+607753125\zeta^{14}\xi^2-4501875\zeta^{14}+560694709575\zeta^{12}\xi^{14} \nonumber \\
    &\quad -1714241754225\zeta^{12}\xi^{12}+2040511317075\zeta^{12}\xi^{10}-1191778867125\zeta^{12}\xi^8+352914547125\zeta^{12}\xi^6 \nonumber \\
    &\quad -49031521875\zeta^{12}\xi^4+2510720625\zeta^{12}\xi^2-19394375\zeta^{12}+824506157475\zeta^{10}\xi^{14}-2558909086965\zeta^{10}\xi^{12} \nonumber \\
    &\quad +3092194963935\zeta^{10}\xi^{10}-1832971953225\zeta^{10}\xi^8+550422482505\zeta^{10}\xi^6-77404535775\zeta^{10}\xi^4 \nonumber \\
    &\quad +3998830325\zeta^{10}\xi^2-31341475\zeta^{10}+615872395215\zeta^8\xi^{14}-1945012278825\zeta^8\xi^{12}+2385803488635\zeta^8\xi^{10} \nonumber \\
    &\quad -1429682265165\zeta^8\xi^8+430957746765\zeta^8\xi^6-60074226555\zeta^8\xi^4+3006587945\zeta^8\xi^2-22101455\zeta^8 \nonumber \\
    &\quad +245626175025\zeta^6\xi^{14}-791771672535\zeta^6\xi^{12}+985062075525\zeta^6\xi^{10}-592389903795\zeta^6\xi^8+175814410163\zeta^6\xi^6 \nonumber \\
    &\quad -23237084805\zeta^6\xi^4+1013428695\zeta^6\xi^2-5099569\zeta^6+50107809885\zeta^4\xi^{14}-165506270475\zeta^4\xi^{12} \nonumber \\
    &\quad +208418460705\zeta^4\xi^{10}-124037312951\zeta^4\xi^8+34794346423\zeta^4\xi^6-3880100385\zeta^4\xi^4+92470027\zeta^4\xi^2+635299\zeta^4 \nonumber \\
    &\quad +4487127225\zeta^2\xi^{14}-15264485775\zeta^2\xi^{12}+19370425773\zeta^2\xi^{10}-11084212811\zeta^2\xi^8+2643918923\zeta^2\xi^6 \nonumber \\
    &\quad -142791597\zeta^2\xi^4-10494065\zeta^2\xi^2+119111\zeta^2+115920525\xi^{14}-408771675\xi^{12}+517889553\xi^{10} \nonumber \\
    &\quad -266212935\xi^8+34318407\xi^6+6821871\xi^4-57765\xi^2+92019) \\
    \mathcal{P}_{2,\text{hMM}}^{(2,2)}&=315(50726300625\zeta^{14}\xi^{14}-153080702775\zeta^{14}\xi^{12}+179545090725\zeta^{14}\xi^{10}-103106185875\zeta^{14}\xi^8 \nonumber \\
    &\quad +29940169875\zeta^{14}\xi^6-4065193125\zeta^{14}\xi^4+202584375\zeta^{14}\xi^2-1500625\zeta^{14}+153080702775\zeta^{12}\xi^{14} \nonumber \\
    &\quad -469360116225\zeta^{12}\xi^{12}+560473711875\zeta^{12}\xi^{10}-328522165125\zeta^{12}\xi^8+97678069125\zeta^{12}\xi^6 \nonumber \\
    &\quad -13633711875\zeta^{12}\xi^4+701850625\zeta^{12}\xi^2-5464375\zeta^{12}+179545090725\zeta^{10}\xi^{14}-560473711875\zeta^{10}\xi^{12} \nonumber \\
    &\quad +681021858825\zeta^{10}\xi^{10}-405723365775\zeta^{10}\xi^8+122347470735\zeta^{10}\xi^6-17254396425\zeta^{10}\xi^4+892054275\zeta^{10}\xi^2 \nonumber \\
    &\quad -7004325\zeta^{10}+103106185875\zeta^8\xi^{14}-328522165125\zeta^8\xi^{12}+405723365775\zeta^8\xi^{10}-243989684505\zeta^8\xi^8 \nonumber \\
    &\quad +73400477465\zeta^8\xi^6-10107342735\zeta^8\xi^4+489496965\zeta^8\xi^2-3340435\zeta^8+29940169875\zeta^6\xi^{14} \nonumber \\
    &\quad -97678069125\zeta^6\xi^{12}+122347470735\zeta^6\xi^{10}-73400477465\zeta^6\xi^8+21340529945\zeta^6\xi^6-2643619215\zeta^6\xi^4 \nonumber \\
    &\quad +92812165\zeta^6\xi^2+3437\zeta^6+4065193125\zeta^4\xi^{14}-13633711875\zeta^4\xi^{12}+17254396425\zeta^4\xi^{10}-10107342735\zeta^4\xi^8 \nonumber \\
    &\quad +2643619215\zeta^4\xi^6-219088905\zeta^4\xi^4-5739261\zeta^4\xi^2+314715\zeta^4+202584375\zeta^2\xi^{14}-701850625\zeta^2\xi^{12} \nonumber \\
    &\quad +892054275\zeta^2\xi^{10} -489496965\zeta^2\xi^8+92812165\zeta^2\xi^6+5739261\zeta^2\xi^4-1692159\zeta^2\xi^2-19255\zeta^2+1500625\xi^{14} \nonumber \\
    &\quad -5464375\xi^{12}+7004325\xi^{10}-3340435\xi^8-3437\xi^6+314715\xi^4+19255\xi^2-30673)
\end{align}

\paragraph{Interaction part $\lambda_\text{int}$:} When a holey Morgan-Morgan disk is superposed with the Schwarzschild black hole, a general expression for the interaction part reads
\begin{align}
    \lambda_\text{hMM,int}^{(m,n)} &= \mathscr{K}^{(m,n)}_{0,\text{hMM}} + \mathscr{K}^{(m,n)}_{1,\text{hMM}} \left\{ \atantwo\left[  \genfrac{}{}{0pt}{}{ b\xi - M\zeta}{ R_-(\zeta, \xi)}\right]-\atantwo\left[ \genfrac{}{}{0pt}{}{ b\xi + M\zeta}{ R_+(\zeta, \xi)} \right] \right\} + \nonumber\\
    & \qquad + \mathscr{K}^{(m,n)}_{2,\text{hMM}}\sum_\mp R_\mp(\zeta, \xi) \left[ \pm\mathscr{P}^{(m,n)}_{0,\text{hMM}}(\pm\zeta, \xi) + \mathscr{P}^{(m,n)}_{1,\text{hMM}}(\pm\zeta, \xi) \arccot(\zeta) \right] \,,
\end{align}
where $\atantwo(\genfrac{}{}{0pt}{}{y}{x}) \equiv \atantwo(y, x)$ denotes 2-argument arcus tangent, $R_\pm(\zeta, \xi)$ comes from the Schwarzschild potential (\ref{eq:RpRmDef}) transformed to the oblate spheroidal coordinates (\ref{eq:CylToOsph}), $\mathscr{K}^{(m,n)}_{j, \text{hMM}}$ stands for constants which depend on the disk parameters, and $\mathscr{P}^{(m,n)}_{j, \text{hMM}}$ are polynomials in $\zeta$ and $\xi$. A few first members read
\begin{align}
    \mathscr{K}_{0,\text{hMM}}^{(1,1)} &= \frac{15 \mathcal{M}}{16 b^5} \left[ -\pi  b^4+2 b^3 M+2 \pi  b^2 M^2+2 (b^2-3 M^2) (b^2+M^2) \arctan\left(\frac{b}{M}\right)+6 b M^3+3 \pi  M^4 \right] \\
    \mathscr{K}_{1,\text{hMM}}^{(1,1)} &= \frac{15 \mathcal{M}}{16 b^5}(b^2-3 M^2) (b^2+M^2) \\
    \mathscr{K}_{2,\text{hMM}}^{(1,1)} &= -\frac{15 \mathcal{M}}{32 b^5} \\
    \mathscr{P}_{0,\text{hMM}}^{(1,1)} &= b(15b^2\zeta^2\xi^3-9b^2\zeta^2\xi+4b^2\xi^3-2b^2\xi-9b\zeta M\xi^2+3b\zeta M+6M^2\xi)\\
    \mathscr{P}_{1,\text{hMM}}^{(1,1)} &= -15b^3\zeta^3\xi^3+9b^3\zeta^3\xi-9b^3\zeta\xi^3+5b^3\zeta\xi+9b^2\zeta^2M\xi^2-3b^2\zeta^2M+3b^2M\xi^2+b^2M-6b\zeta M^2\xi+6M^3\\
    \mathscr{K}_{0,\text{hMM}}^{(1,2)} &= \frac{35 \mathcal{M}}{96 b^7} \left[ 6 b^5 M+44 b^3 M^3-3 \pi  (b^2-5 M^2) (b^2+M^2)^2+6 (b^2-5 M^2)(b^2+M^2)^2 \arctan\left(\frac{b}{M}\right)+30 b M^5 \right] \\
    \mathscr{K}_{1,\text{hMM}}^{(1,2)} &= (b^2-5 M^2) (b^2+M^2)^2 \\
    \mathscr{K}_{2,\text{hMM}}^{(1,2)} &= -\frac{35 \mathcal{M}}{768 b^7} \\
    \mathscr{P}_{0,\text{hMM}}^{(1,2)}&=b(945b^4\zeta^4\xi^5-1050b^4\zeta^4\xi^3+225b^4\zeta^4\xi+735b^4\zeta^2\xi^5-610b^4\zeta^2\xi^3+51b^4\zeta^2\xi+64b^4\xi^5-16b^4\xi^3-24b^4\xi \nonumber\\
    & \quad -525b^3\zeta^3M\xi^4+450b^3\zeta^3M\xi^2-45b^3\zeta^3M-275b^3\zeta M\xi^4+66b^3\zeta M\xi^2+33b^3\zeta M+300b^2\zeta^2M^2\xi^3 \nonumber\\
    & \quad -180b^2\zeta^2M^2\xi+80b^2M^2\xi^3+96b^2M^2\xi-180b\zeta M^3\xi^2+60b\zeta M^3+120M^4\xi) \\
    \mathscr{P}_{1,\text{hMM}}^{(1,2)} &=-3(315b^5\zeta^5\xi^5-350b^5\zeta^5\xi^3+75b^5\zeta^5\xi+350b^5\zeta^3\xi^5-320b^5\zeta^3\xi^3+42b^5\zeta^3\xi+75b^5\zeta\xi^5-42b^5\zeta\xi^3-9b^5\zeta\xi \nonumber \\
    &\quad -175b^4\zeta^4M\xi^4+150b^4\zeta^4M\xi^2-15b^4\zeta^4M-150b^4\zeta^2M\xi^4+72b^4\zeta^2M\xi^2+6b^4\zeta^2M-15b^4M\xi^4-6b^4M\xi^2 \nonumber \\
    &\quad -3b^4M+100b^3\zeta^3M^2\xi^3-60b^3\zeta^3M^2\xi+60b^3\zeta M^2\xi^3+12b^3\zeta M^2\xi-60b^2\zeta^2M^3\xi^2+20b^2\zeta^2M^3 \nonumber \\
    &\quad -20b^2M^3\xi^2-52b^2M^3+40b\zeta M^4\xi-40M^5)\\
    \mathscr{K}_{0,\text{hMM}}^{(2,1)} &=\frac{35\mathcal{M}}{128b^7}\left[6b^5M-8b^3M^3-3\pi(b^2+M^2)(b^4-2b^2M^2+5M^4)-30bM^5+\right. \nonumber \\
    &\quad \left. +6 (b^2+M^2)(b^4-2 b^2 M^2+5 M^4) \arctan\left(\frac{b}{M}\right) \right] \\
    \mathscr{K}_{1,\text{hMM}}^{(2,1)} &= \frac{105 \mathcal{M}}{128 b^7} (b^2+M^2) (b^4-2 b^2 M^2+5 M^4) \\
    \mathscr{K}_{2,\text{hMM}}^{(2,1)} &= \frac{35 \mathcal{M}}{1024 b^7} \\
    \mathscr{P}_{0,\text{hMM}}^{(2,1)} &= b(945b^4\zeta^4\xi^5-1050b^4\zeta^4\xi^3+225b^4\zeta^4\xi+735b^4\zeta^2\xi^5-970b^4\zeta^2\xi^3+267b^4\zeta^2\xi+64b^4\xi^5-112b^4\xi^3+24b^4\xi \nonumber \\
    &\quad -525b^3\zeta^3M\xi^4+450b^3\zeta^3M\xi^2-45b^3\zeta^3M-275b^3\zeta M\xi^4+282b^3\zeta M\xi^2-39b^3\zeta M+300b^2\zeta^2M^2\xi^3 \nonumber \\
    &\quad -180b^2\zeta^2M^2\xi+80b^2M^2\xi^3-48b^2M^2\xi-180b\zeta M^3\xi^2+60b\zeta M^3+120M^4\xi)\\
    \mathscr{P}_{1,\text{hMM}}^{(2,1)} &= -3(315b^5\zeta^5\xi^5-350b^5\zeta^5\xi^3+75b^5\zeta^5\xi+350b^5\zeta^3\xi^5-440b^5\zeta^3\xi^3+114b^5\zeta^3\xi+75b^5\zeta\xi^5-114b^5\zeta\xi^3 \nonumber \\
    &\quad +31b^5\zeta\xi-175b^4\zeta^4M\xi^4+150b^4\zeta^4M\xi^2-15b^4\zeta^4M-150b^4\zeta^2M\xi^4+144b^4\zeta^2M\xi^2-18b^4\zeta^2M \nonumber \\
    &\quad -15b^4M\xi^4+18b^4M\xi^2+5b^4M+100b^3\zeta^3M^2\xi^3-60b^3\zeta^3M^2\xi+60b^3\zeta M^2\xi^3-36b^3\zeta M^2\xi-60b^2\zeta^2M^3\xi^2 \nonumber \\
    &\quad +20b^2\zeta^2M^3-20b^2M^3\xi^2-4b^2M^3+40b\zeta M^4\xi-40M^5) \\
    \mathscr{K}_{0,\text{hMM}}^{(2,2)} &= \frac{105 \mathcal{M}}{1024 b^9} \left[ 18 b^7 M-30 b^5 M^3-290 b^3 M^5-3 \pi  (b^2+M^2)^2 (3 b^4-10 b^2 M^2+35 M^4)+ \right. \nonumber \\
    &\qquad\qquad \left. +6(b^2+M^2)^2 (3 b^4-10 b^2 M^2+35 M^4) \arctan\left(\frac{b}{M}\right)-210 b M^7 \right] \\
    \mathscr{K}_{1,\text{hMM}}^{(2,2)} &= \frac{105 \mathcal{M}}{1024 b^9} (b^2+M^2)^2 (3 b^4-10 b^2 M^2+35 M^4) \\
    \mathscr{K}_{2,\text{hMM}}^{(2,2)} &= \frac{105 \mathcal{M}}{16384 b^9} \\
    \mathscr{P}_{0,\text{hMM}}^{(2,2)} &= b(45045b^6\zeta^6\xi^7-72765b^6\zeta^6\xi^5+33075b^6\zeta^6\xi^3-3675b^6\zeta^6\xi+57750b^6\zeta^4\xi^7-91980b^6\zeta^4\xi^5+40950b^6\zeta^4\xi^3 \nonumber \\
    &\quad -4400b^6\zeta^4\xi+17829b^6\zeta^2\xi^7-27783b^6\zeta^2\xi^5+10575b^6\zeta^2\xi^3-381b^6\zeta^2\xi+768b^6\xi^7-1152b^6\xi^5+96b^6\xi^3 \nonumber \\
    &\quad +144b^6\xi-24255b^5\zeta^5M\xi^6+33075b^5\zeta^5M\xi^4-11025b^5\zeta^5M\xi^2+525b^5\zeta^5M-24990b^5\zeta^3M\xi^6 \nonumber \\
    &\quad +31500b^5\zeta^3M\xi^4-9150b^5\zeta^3M\xi^2+320b^5\zeta^3M-4851b^5\zeta M\xi^6+5265b^5\zeta M\xi^4-441b^5\zeta M\xi^2-213b^5\zeta M \nonumber \\
    &\quad +13230b^4\zeta^4M^2\xi^5-14700b^4\zeta^4M^2\xi^3+3150b^4\zeta^4M^2\xi+10290b^4\zeta^2M^2\xi^5-8900b^4\zeta^2M^2\xi^3+930b^4\zeta^2M^2\xi \nonumber \\
    &\quad +896b^4M^2\xi^5-320b^4M^2\xi^3-336b^4M^2\xi-7350b^3\zeta^3M^3\xi^4+6300b^3\zeta^3M^3\xi^2-630b^3\zeta^3M^3 \nonumber \\
    &\quad -3850b^3\zeta M^3\xi^4+1140b^3\zeta M^3\xi^2+390b^3\zeta M^3+4200b^2\zeta^2M^4\xi^3-2520b^2\zeta^2M^4\xi+1120b^2M^4\xi^3 \nonumber \\
    &\quad +1200b^2M^4\xi-2520b\zeta M^5\xi^2+840b\zeta M^5+1680M^6\xi) \\
    \mathscr{P}_{1,\text{hMM}}^{(2,2)} &= -3(15015b^7\zeta^7\xi^7-24255b^7\zeta^7\xi^5+11025b^7\zeta^7\xi^3-1225b^7\zeta^7\xi+24255b^7\zeta^5\xi^7-38745b^7\zeta^5\xi^5+17325b^7\zeta^5\xi^3 \nonumber \\
    &\quad -1875b^7\zeta^5\xi+11025b^7\zeta^3\xi^7-17325b^7\zeta^3\xi^5+7095b^7\zeta^3\xi^3-507b^7\zeta^3\xi+1225b^7\zeta\xi^7-1875b^7\zeta\xi^5+507b^7\zeta\xi^3 \nonumber \\
    &\quad +79b^7\zeta\xi-8085b^6\zeta^6M\xi^6+11025b^6\zeta^6M\xi^4-3675b^6\zeta^6M\xi^2+175b^6\zeta^6M-11025b^6\zeta^4M\xi^6 \nonumber \\
    &\quad +14175b^6\zeta^4M\xi^4-4275b^6\zeta^4M\xi^2+165b^6\zeta^4M-3675b^6\zeta^2M\xi^6+4275b^6\zeta^2M\xi^4-837b^6\zeta^2M\xi^2 \nonumber \\
    &\quad -51b^6\zeta^2M-175b^6M\xi^6+165b^6M\xi^4+51b^6M\xi^2+23b^6M+4410b^5\zeta^5M^2\xi^5-4900b^5\zeta^5M^2\xi^3 \nonumber \\
    &\quad +1050b^5\zeta^5M^2\xi+4900b^5\zeta^3M^2\xi^5-4600b^5\zeta^3M^2\xi^3+660b^5\zeta^3M^2\xi+1050b^5\zeta M^2\xi^5-660b^5\zeta M^2\xi^3 \nonumber \\
    &\quad -102b^5\zeta M^2\xi-2450b^4\zeta^4M^3\xi^4+2100b^4\zeta^4M^3\xi^2-210b^4\zeta^4M^3-2100b^4\zeta^2M^3\xi^4+1080b^4\zeta^2M^3\xi^2 \nonumber \\
    &\quad +60b^4\zeta^2M^3-210b^4M^3\xi^4-60b^4M^3\xi^2-18b^4M^3+1400b^3\zeta^3M^4\xi^3-840b^3\zeta^3M^4\xi+840b^3\zeta M^4\xi^3 \nonumber \\
    &\quad +120b^3\zeta M^4\xi-840b^2\zeta^2M^5\xi^2+280b^2\zeta^2M^5-280b^2M^5\xi^2-680b^2M^5+560b\zeta M^6\xi-560M^7)
\end{align}

\subsection{Inverted holey Morgan-Morgan disks}

\paragraph{Quadratic $\lambda_\text{disk}$:} For the inverted holey Morgan-Morgan disks, a general expression for the second metric function (\ref{eq:lambdaihMM}) reads
\begin{align}
    \lambda_\text{ihMM}^{(m,n)} &= \mathcal{C}_{\text{ihMM}}^{(m,n)}\left[ \frac{2 (m + n)!}{\pi \left(\frac{1}{2}\right)_m \left(\frac{1}{2}\right)_n} \right]^2 \frac{\mathcal{M}^2}{b^2} \mathcal{K} \left\{(\xi^2 - 1) \left[ \mathcal{P}_{0, \text{ihMM}}^{(m,n)} + 2\zeta \mathcal{P}_{1, \text{ihMM}}^{(m,n)} \arccot(\zeta) + (\zeta^2 + 1) \mathcal{P}_{2, \text{ihMM}}^{(m,n)} \arccot^{2}(\zeta) \right] \right\} \,,
\end{align}
where $\mathcal{P}_{j, \text{ihMM}}^{(m,n)}$ are polynomials in $(\zeta, \xi)$, and $\mathcal{C}_{\text{ihMM}}^{(m,n)}$ are numerical constants. For the first few members, we have explicitly
\begin{align}
    \mathcal{C}_{\text{ihMM}}^{(0,1)} &= \frac{1}{96} \\
    \mathcal{P}_{0, \text{ihMM}}^{(0,1)} &= 225 \zeta ^4 \xi ^4-126 \zeta ^4 \xi ^2+9 \zeta ^4+201 \zeta ^2 \xi ^4-96 \zeta ^2 \xi ^2+3 \zeta ^2+16 \xi^4-8 \xi ^2-8 \\
    \mathcal{P}_{1, \text{ihMM}}^{(0,1)} &= -3 (75 \zeta ^4 \xi ^4-42 \zeta ^4 \xi ^2+3 \zeta ^4+92 \zeta ^2 \xi ^4-46 \zeta ^2 \xi ^2+2 \zeta ^2+21\xi ^4-8 \xi ^2-1) \\
    \mathcal{P}_{2, \text{ihMM}}^{(0,1)} &= 3 (75 \zeta ^4 \xi ^4-42 \zeta ^4 \xi ^2+3 \zeta ^4+42 \zeta ^2 \xi ^4-18 \zeta ^2 \xi ^2+3 \xi^4+1) \\
    \mathcal{C}_{\text{ihMM}}^{(0,2)} &= \frac{1}{40960} \\
    \mathcal{P}_{0,\text{ihMM}}^{(0,2)} &= 893025\zeta^8\xi^8-1367100\zeta^8\xi^6+633150\zeta^8\xi^4-89100\zeta^8\xi^2+2025\zeta^8+1664775\zeta^6\xi^8-2238600\zeta^6\xi^6 \nonumber \\
    &\quad +856350\zeta^6\xi^4-86400\zeta^6\xi^2+675\zeta^6+949935\zeta^4\xi^8-1026940\zeta^4\xi^6+262410\zeta^4\xi^4-6300\zeta^4\xi^2-225\zeta^4 \nonumber \\
    &\quad +169705\zeta^2\xi^8-120320\zeta^2\xi^6+3930\zeta^2\xi^4+360\zeta^2\xi^2+405\zeta^2+4096\xi^8-1024\xi^6-1024\xi^4-1024\xi^2 \nonumber \\
    &\quad -1024 \\
    \mathcal{P}_{1,\text{ihMM}}^{(0,2)}&=-15(59535\zeta^8\xi^8-91140\zeta^8\xi^6+42210\zeta^8\xi^4-5940\zeta^8\xi^2+135\zeta^8+130830\zeta^6\xi^8-179620\zeta^6\xi^6+71160\zeta^6\xi^4 \nonumber \\
    &\quad -7740\zeta^6\xi^2+90\zeta^6+95032\zeta^4\xi^8-110108\zeta^4\xi^6+32772\zeta^4\xi^4-1812\zeta^4\xi^2-12\zeta^4+25330\zeta^2\xi^8-21820\zeta^2\xi^6 \nonumber \\
    &\quad +3112\zeta^2\xi^4+156\zeta^2\xi^2+6\zeta^2+1785\xi^8-768\xi^6-134\xi^4-24\xi^2-27) \\
    \mathcal{P}_{2,\text{ihMM}}^{(0,2)}&=45(19845\zeta^8\xi^8-30380\zeta^8\xi^6+14070\zeta^8\xi^4-1980\zeta^8\xi^2+45\zeta^8+30380\zeta^6\xi^8-39620\zeta^6\xi^6+14340\zeta^6\xi^4 \nonumber \\
    &\quad -1260\zeta^6\xi^2+14070\zeta^4\xi^8-14340\zeta^4\xi^6+3240\zeta^4\xi^4-28\zeta^4\xi^2+2\zeta^4+1980\zeta^2\xi^8-1260\zeta^2\xi^6 \nonumber \\
    &\quad +28\zeta^2\xi^4+28\zeta^2\xi^2-8\zeta^2+45\xi^8+2\xi^4+8\xi^2+9) \\
    \mathcal{C}_{\text{ihMM}}^{(1,1)} &= \frac{1}{40960} \\
    \mathcal{P}_{0,\text{ihMM}}^{(1,1)} &= 893025\zeta^8\xi^8-1367100\zeta^8\xi^6+633150\zeta^8\xi^4-89100\zeta^8\xi^2+2025\zeta^8+1664775\zeta^6\xi^8-2805600\zeta^6\xi^6 \nonumber \\
    &\quad +1448550\zeta^6\xi^4-232200\zeta^6\xi^2+6075\zeta^6+949935\zeta^4\xi^8-1808140\zeta^4\xi^6+1050810\zeta^4\xi^4-187980\zeta^4\xi^2 \nonumber \\
    &\quad +5295\zeta^4+169705\zeta^2\xi^8-375320\zeta^2\xi^6+250850\zeta^2\xi^4-45280\zeta^2\xi^2+365\zeta^2+4096\xi^8-11264\xi^6+9216\xi^4 \nonumber \\
    &\quad -1024\xi^2-1024 \\
    \mathcal{P}_{1,\text{ihMM}}^{(1,1)}&=-5(178605\zeta^8\xi^8-273420\zeta^8\xi^6+126630\zeta^8\xi^4-17820\zeta^8\xi^2+405\zeta^8+392490\zeta^6\xi^8-652260\zeta^6\xi^6 \nonumber \\
    &\quad +331920\zeta^6\xi^4-52380\zeta^6\xi^2+1350\zeta^6+285096\zeta^4\xi^8-524364\zeta^4\xi^6+295476\zeta^4\xi^4-51492\zeta^4\xi^2+1428\zeta^4 \nonumber \\
    &\quad +75990\zeta^2\xi^8-158460\zeta^2\xi^6+99856\zeta^2\xi^4-18436\zeta^2\xi^2+410\zeta^2+5355\xi^8-13128\xi^6+9350\xi^4-1568\xi^2-73) \\
    \mathcal{P}_{2,\text{ihMM}}^{(1,1)}&=5(178605\zeta^8\xi^8-273420\zeta^8\xi^6+126630\zeta^8\xi^4-17820\zeta^8\xi^2+405\zeta^8+273420\zeta^6\xi^8-469980\zeta^6\xi^6 \nonumber \\
    &\quad +247500\zeta^6\xi^4-40500\zeta^6\xi^2+1080\zeta^6+126630\zeta^4\xi^8-247500\zeta^4\xi^6+147360\zeta^4\xi^4-26868\zeta^4\xi^2+762\zeta^4 \nonumber \\
    &\quad +17820\zeta^2\xi^8-40500\zeta^2\xi^6+26868\zeta^2\xi^4-5052\zeta^2\xi^2+96\zeta^2+405\xi^8-1080\xi^6+762\xi^4-96\xi^2+73) \\
    \mathcal{C}_{\text{ihMM}}^{(1,2)} &= \frac{1}{41287680} \\
    \mathcal{P}_{0,\text{ihMM}}^{(1,2)} &= 10145260125\zeta^{12}\xi^{12}-25573259250\zeta^{12}\xi^{10}+23821441875\zeta^{12}\xi^8-10044877500\zeta^{12}\xi^6+1872871875\zeta^{12}\xi^4 \nonumber \\
    &\quad -124031250\zeta^{12}\xi^2+1378125\zeta^{12}+28955012625\zeta^{10}\xi^{12}-72801382500\zeta^{10}\xi^{10}+67610426625\zeta^{10}\xi^8 \nonumber \\
    &\quad -28406574000\zeta^{10}\xi^6+5272981875\zeta^{10}\xi^4-347287500\zeta^{10}\xi^2+3766875\zeta^{10}+30767710050\zeta^8\xi^{12} \nonumber \\
    &\quad -77117537700\zeta^8\xi^{10}+70975576350\zeta^8\xi^8-29263495800\zeta^8\xi^6+5239763550\zeta^8\xi^4-321545700\zeta^8\xi^2 \nonumber \\
    &\quad +2934050\zeta^8+14941303650\zeta^6\xi^{12}-37303988400\zeta^6\xi^{10}+33648082650\zeta^6\xi^8-13213628400\zeta^6\xi^6 \nonumber \\
    &\quad +2130600150\zeta^6\xi^4-102093600\zeta^6\xi^2+344750\zeta^6+3245151105\zeta^4\xi^{12}-8062033770\zeta^4\xi^{10} \nonumber \\
    &\quad +6975883215\zeta^4\xi^8-2440446540\zeta^4\xi^6+290714655\zeta^4\xi^4-3789450\zeta^4\xi^2-22575\zeta^4+254587725\zeta^2\xi^{12} \nonumber \\
    &\quad -628428780\zeta^2\xi^{10}+500078565\zeta^2\xi^8-128429280\zeta^2\xi^6+2173815\zeta^2\xi^4+41580\zeta^2\xi^2+57015\zeta^2 \nonumber \\
    &\quad +2949120\xi^{12}-7372800\xi^{10}+5013504\xi^8-147456\xi^6-147456\xi^4-147456\xi^2-147456 \\
    \mathcal{P}_{1,\text{ihMM}}^{(1,2)}&=-105(96621525\zeta^{12}\xi^{12}-243554850\zeta^{12}\xi^{10}+226870875\zeta^{12}\xi^8-95665500\zeta^{12}\xi^6+17836875\zeta^{12}\xi^4 \nonumber \\
    &\quad -1181250\zeta^{12}\xi^2+13125\zeta^{12}+307969200\zeta^{10}\xi^{12}-774531450\zeta^{10}\xi^{10}+719532450\zeta^{10}\xi^8-302427300\zeta^{10}\xi^6 \nonumber \\
    &\quad +56164500\zeta^{10}\xi^4-3701250\zeta^{10}\xi^2+40250\zeta^{10}+376357905\zeta^8\xi^{12}-943918920\zeta^8\xi^{10}+870427845\zeta^8\xi^8 \nonumber \\
    &\quad -360375960\zeta^8\xi^6+65056635\zeta^8\xi^4-4059840\zeta^8\xi^2+38735\zeta^8+219960000\zeta^6\xi^{12}-549802980\zeta^6\xi^{10} \nonumber \\
    &\quad +499104180\zeta^6\xi^8-199150440\zeta^6\xi^6+33292200\zeta^6\xi^4-1754100\zeta^6\xi^2+10020\zeta^6+62214495\zeta^4\xi^{12} \nonumber \\
    &\quad -154850850\zeta^4\xi^{10}+136301949\zeta^4\xi^8-50147564\zeta^4\xi^6+6902553\zeta^4\xi^4-200178\zeta^4\xi^2-1013\zeta^4 \nonumber \\
    &\quad +7501200\zeta^2\xi^{12}-18569250\zeta^2\xi^{10}+15388458\zeta^2\xi^8-4689140\zeta^2\xi^6+334788\zeta^2\xi^4+14454\zeta^2\xi^2+34\zeta^2 \nonumber \\
    &\quad +255675\xi^{12}-628980\xi^{10}+459747\xi^8-76320\xi^6-9951\xi^4-396\xi^2-543) \\
    \mathcal{P}_{2,\text{ihMM}}^{(1,2)}&=315(32207175\zeta^{12}\xi^{12}-81184950\zeta^{12}\xi^{10}+75623625\zeta^{12}\xi^8-31888500\zeta^{12}\xi^6+5945625\zeta^{12}\xi^4-393750\zeta^{12}\xi^2 \nonumber \\
    &\quad +4375\zeta^{12}+81184950\zeta^{10}\xi^{12}-204053850\zeta^{10}\xi^{10}+189428400\zeta^{10}\xi^8-79550100\zeta^{10}\xi^6+14757750\zeta^{10}\xi^4 \nonumber \\
    &\quad -971250\zeta^{10}\xi^2+10500\zeta^{10}+75623625\zeta^8\xi^{12}-189428400\zeta^8\xi^{10}+173940165\zeta^8\xi^8-71343720\zeta^8\xi^6 \nonumber \\
    &\quad +12639795\zeta^8\xi^4-758280\zeta^8\xi^2+6495\zeta^8+31888500\zeta^6\xi^{12}-79550100\zeta^6\xi^{10}+71343720\zeta^6\xi^8 \nonumber \\
    &\quad -27605480\zeta^6\xi^6+4298820\zeta^6\xi^4-186180\zeta^6\xi^2+160\zeta^6+5945625\zeta^4\xi^{12}-14757750\zeta^4\xi^{10}+12639795\zeta^4\xi^8 \nonumber \\
    &\quad -4298820\zeta^4\xi^6+467775\zeta^4\xi^4+1338\zeta^4\xi^2-267\zeta^4+393750\zeta^2\xi^{12}-971250\zeta^2\xi^{10}+758280\zeta^2\xi^8 \nonumber \\
    &\quad -186180\zeta^2\xi^6-1338\zeta^2\xi^4+2262\zeta^2\xi^2-132\zeta^2+4375\xi^{12}-10500\xi^{10}+6495\xi^8-160\xi^6-267\xi^4+132\xi^2 \nonumber \\
    &\quad +181)
\end{align}

\paragraph{Interaction part $\lambda_\text{int}$:} A general expression for the interaction part of $\lambda$ for the superposition of the inverted holey Morgan-Morgan disks with the Schwarzschild black hole (\ref{eq:lambdaintihMM}) is
\begin{align}
    \lambda_\text{ihMM,int}^{(m,n)} &= \mathscr{K}^{(m,n)}_{1,\text{ihMM}} \left\{ \atantwo\left[  \genfrac{}{}{0pt}{}{ b\xi - M\zeta}{ R_-(\zeta, \xi)}\right]+\atantwo\left[ \genfrac{}{}{0pt}{}{ b\xi + M\zeta}{ R_+(\zeta, \xi)} \right] \right\} + \nonumber\\
    & \qquad + \mathscr{K}^{(m,n)}_{2,\text{ihMM}}\sum_\mp R_\mp(\zeta, \xi) \, \mathcal{K} \left\{\sqrt{1 + \zeta^2 - \xi^2}  \left[ \pm \mathscr{P}^{(m,n)}_{0,\text{ihMM}}(\pm\zeta, \xi) + \mathscr{P}^{(m,n)}_{1,\text{ihMM}}(\pm\zeta, \xi) \arccot(\zeta) \right] \right\} \,,
\end{align}
with the same notation conventions as in (\ref{eq:app:lambdahMM}), namely $\mathscr{K}^{(m,n)}_{j, \text{ihMM}}$ are constants and $\mathscr{P}^{(m,n)}_{j, \text{ihMM}}$ are polynomials in $\zeta$ and~$\xi$. A few first members read explicitly
\begin{align}
    \mathscr{K}_{1,\text{ihMM}}^{(0,1)} &= \frac{2 \mathcal{M} \left(b^2+M^2\right)}{\pi  M^3} \\
    \mathscr{K}_{2,\text{ihMM}}^{(0,1)} &= \frac{\mathcal{M}}{\pi  b M^2} \\
    \mathscr{P}_{0,\text{ihMM}}^{(0,1)} &= -2 b \xi +3 \zeta  M \xi ^2-\zeta  M \\
    \mathscr{P}_{1,\text{ihMM}}^{(0,1)} &= \frac{1}{M}(-2 b^2+2 b \zeta  M \xi -3 \zeta ^2 M^2 \xi ^2+\zeta ^2 M^2-M^2 \xi ^2-M^2) \\
    \mathscr{K}_{1,\text{ihMM}}^{(0,2)} &= \frac{2 \mathcal{M} \left(b^2+M^2\right)^2}{\pi  M^5} \\
    \mathscr{K}_{2,\text{ihMM}}^{(0,2)} &= \frac{\mathcal{M}}{12 \pi  b M^4} \\
    \mathscr{P}_{0,\text{ihMM}}^{(0,2)} &= -24b^3\xi+36b^2\zeta M\xi^2-12b^2\zeta M-60b\zeta^2M^2\xi^3+36b\zeta^2M^2\xi-16bM^2\xi^3-24bM^2\xi+105\zeta^3M^3\xi^4 \nonumber \\
    &\quad -90\zeta^3M^3\xi^2+9\zeta^3M^3+55\zeta M^3\xi^4-6\zeta M^3\xi^2-9\zeta M^3 \\
    \mathscr{P}_{1,\text{ihMM}}^{(0,2)} &= \frac{-3}{M}(8b^4-8b^3\zeta M\xi+12b^2\zeta^2M^2\xi^2-4b^2\zeta^2M^2+4b^2M^2\xi^2+12b^2M^2-20b\zeta^3M^3\xi^3+12b\zeta^3M^3\xi \nonumber \\
    &\quad -12b\zeta M^3\xi^3-4b\zeta M^3\xi+35\zeta^4M^4\xi^4-30\zeta^4M^4\xi^2+3\zeta^4M^4+30\zeta^2M^4\xi^4-12\zeta^2M^4\xi^2-2\zeta^2M^4+3M^4\xi^4 \nonumber \\
    &\quad +2M^4\xi^2+3M^4) \\
    \mathscr{K}_{1,\text{ihMM}}^{(1,1)} &= \frac{2 \mathcal{M} \left(-3 b^4-2 b^2 M^2+M^4\right)}{\pi  M^5} \\
    \mathscr{K}_{2,\text{ihMM}}^{(1,1)} &= \frac{\mathcal{M}}{4 \pi  b M^4} \\
    \mathscr{P}_{0,\text{ihMM}}^{(1,1)} &= 24b^3\xi -36b^2\zeta M\xi^2+12b^2\zeta M+60b\zeta^2M^2\xi^3-36b\zeta^2M^2\xi+16bM^2\xi^3-8bM^2\xi-105\zeta^3M^3\xi^4+90\zeta^3M^3\xi^2 \nonumber \\
    &\quad -9\zeta^3M^3-55\zeta M^3\xi^4+54\zeta M^3\xi^2-7\zeta M^3 \\
    \mathscr{P}_{1,\text{ihMM}}^{(1,1)} &= \frac{1}{M} (24b^4-24b^3\zeta M\xi+36b^2\zeta^2M^2\xi^2-12b^2\zeta^2M^2+12b^2M^2\xi^2+4b^2M^2-60b\zeta^3M^3\xi^3+36b\zeta^3M^3\xi \nonumber \\
    &\quad -36b\zeta M^3\xi^3+20b\zeta M^3\xi+105\zeta^4M^4\xi^4-90\zeta^4M^4\xi^2+9\zeta^4M^4+90\zeta^2M^4\xi^4-84\zeta^2M^4\xi^2+10\zeta^2M^4 \nonumber \\
    &\quad +9M^4\xi^4-10M^4\xi^2-7M^4) \\
    \mathscr{K}_{1,\text{ihMM}}^{(1,2)} &= \frac{2 \mathcal{M} \left(M^2-5 b^2\right) \left(b^2+M^2\right)^2}{\pi  M^7} \\
    \mathscr{K}_{2,\text{ihMM}}^{(1,2)} &= \frac{\mathcal{M}}{24 \pi  b M^6}\\
    \mathscr{P}_{0,\text{ihMM}}^{(1,2)} &= 240b^5\xi-360b^4\zeta M\xi^2+120b^4\zeta M+600b^3\zeta^2M^2\xi^3-360b^3\zeta^2M^2\xi+160b^3M^2\xi^3+192b^3M^2\xi \nonumber \\
    &\quad -1050b^2\zeta^3M^3\xi^4+900b^2\zeta^3M^3\xi^2-90b^2\zeta^3M^3-550b^2\zeta M^3\xi^4+132b^2\zeta M^3\xi^2+66b^2\zeta M^3+1890b\zeta^4M^4\xi^5 \nonumber \\
    &\quad -2100b\zeta^4M^4\xi^3+450b\zeta^4M^4\xi+1470b\zeta^2M^4\xi^5-1220b\zeta^2M^4\xi^3+102b\zeta^2M^4\xi+128bM^4\xi^5-32bM^4\xi^3 \nonumber \\
    &\quad -48bM^4\xi-3465\zeta^5M^5\xi^6+4725\zeta^5M^5\xi^4-1575\zeta^5M^5\xi^2+75\zeta^5M^5-3570\zeta^3M^5\xi^6+4410\zeta^3M^5\xi^4 \nonumber \\
    &\quad -1230\zeta^3M^5\xi^2+38\zeta^3M^5-693\zeta M^5\xi^6+705\zeta M^5\xi^4-27\zeta M^5\xi^2-33\zeta M^5\\
    \mathscr{P}_{1,\text{ihMM}}^{(1,2)} &= \frac{3}{M} (80b^6-80b^5\zeta M\xi+120b^4\zeta^2M^2\xi^2-40b^4\zeta^2M^2+40b^4M^2\xi^2+104b^4M^2-200b^3\zeta^3M^3\xi^3+120b^3\zeta^3M^3\xi \nonumber \\
    &\quad -120b^3\zeta M^3\xi^3-24b^3\zeta M^3\xi+350b^2\zeta^4M^4\xi^4-300b^2\zeta^4M^4\xi^2+30b^2\zeta^4M^4+300b^2\zeta^2M^4\xi^4-144b^2\zeta^2M^4\xi^2 \nonumber \\
    &\quad -12b^2\zeta^2M^4+30b^2M^4\xi^4+12b^2M^4\xi^2+6b^2M^4-630b\zeta^5M^5\xi^5+700b\zeta^5M^5\xi^3-150b\zeta^5M^5\xi \nonumber \\
    &\quad -700b\zeta^3M^5\xi^5+640b\zeta^3M^5\xi^3-84b\zeta^3M^5\xi-150b\zeta M^5\xi^5+84b\zeta M^5\xi^3+18b\zeta M^5\xi+1155\zeta^6M^6\xi^6 \nonumber \\
    &\quad -1575\zeta^6M^6\xi^4+525\zeta^6M^6\xi^2-25\zeta^6M^6+1575\zeta^4M^6\xi^6-1995\zeta^4M^6\xi^4+585\zeta^4M^6\xi^2-21\zeta^4M^6 \nonumber \\
    &\quad +525\zeta^2M^6\xi^6-585\zeta^2M^6\xi^4+99\zeta^2M^6\xi^2+9\zeta^2M^6+25M^6\xi^6-21M^6\xi^4-9M^6\xi^2-11M^6)
\end{align}

\endgroup

\twocolumngrid
\nocite{*}

\bibliography{kokos}

\end{document}